\documentclass[pdflatex,sn-mathphys-num]{sn-jnl}

\usepackage{graphicx}%
\usepackage{multirow}%
\usepackage{amsmath,amssymb,amsfonts}%
\usepackage{amsthm}%
\usepackage{mathrsfs}%
\usepackage[title]{appendix}%
\usepackage{xcolor}%
\usepackage{textcomp}%
\usepackage{manyfoot}%
\usepackage{booktabs}%
\usepackage{algorithm}%
\usepackage{algorithmicx}%
\usepackage{algpseudocode}%
\usepackage{listings}%
\usepackage{stmaryrd}

\theoremstyle{thmstyleone}%
\theoremstyle{thmstyletwo}%

\theoremstyle{thmstylethree}%

\begin{document}

\title[Article Title]{Operational reconstruction of Feynman rules for quantum amplitudes
via composition algebras}

\author*[1]{\fnm{Jens} \sur{K\"oplinger}}\email{jenskoeplinger@gmail.com}

\author[2]{\fnm{Michael} \sur{Habeck}}\email{michael.habeck@uni-jena.de}

\author[3]{\fnm{Philip} \sur{Goyal}}\email{pgoyal@albany.edu}

\affil*[1]{\orgaddress{\street{105 E Avondale Dr}, \city{Greensboro}, \postcode{27403}, \state{NC}, \country{USA}}}

\affil[2]{\orgdiv{Department of Computer Science}, \orgname{University of Jena}, \orgaddress{\country{Germany}}}

\affil[3]{\orgdiv{Department of Physics}, \orgname{University at Albany (SUNY)}, \orgaddress{\state{NY}, \country{USA}}}

\abstract{This article explores an operational model for transition amplitudes between measurements proposed by Goyal et al.~within the quantum reconstruction program. To classify suitable amplitude algebras, we distinguish mathematical axioms, physical choices, and their consequences. This leads to several improvements on the published work: Our coordinate-independent approach requires no two-dimensional amplitudes a priori. All scalar field and vector space axioms are traced from model axioms and observer choices, including additive and multiplicative units and inverses. Existing mathematical characterizations identify allowable amplitude algebras as the real associative composition algebras, namely the complex numbers and the quaternions, as well as their split forms. Observed probabilities are quadratic in amplitudes, akin to the Born rule. We examine selected implications of the proposed axioms, reformulate observer questions, and highlight the broad applicability of our framework to subsequent discovery.}


\maketitle

\section{Introduction}

\subsection{Context}

The Dirac--von Neumann axioms of quantum theory possess a rich mathematical
structure. However, unlike the theories of physics which preceded
it (such as Newtonian mechanics and Maxwell's theory of electromagnetism),
it is difficult to account for this structure by pointing to specific
physical principles, facts about the physical world, or theoretical
choices in how we represent that world symbolically. This predicament
is the legacy of the rather convoluted process by which the quantum
formalism was historically obtained, a process that heavily relied
on inspired mathematical guesswork, as well as new physical principles
and phenomena.

Since the creation of quantum formalism a century ago, there have
been many attempts to rectify this situation (e.g.~\cite{BvN1936,Mackey1957,EP1963,Fivel1994}).
However, these attempts have had comparatively little impact, in part
due to the rather abstract nature of some of their assumptions. The
rise of quantum information in the 1980s led to the hypothesis that
quantum theory might be derivable in terms of information-theoretic
principles \cite{Rovelli1996,Fuchs2003}, and to the conviction such
a derivation could be achieved without recourse to abstract mathematical
assumptions that lack clear physical justification \cite{Hardy2001,Grin2003,Grin2007,Goyal2023}.
Following promising early information-inspired reconstructive attempts
(e.g.~\cite{Wootters1980}), one of the first successful reconstructions
of the finite-dimensional quantum axioms was carried out by Lucien
Hardy \cite{Hardy2001,Hardy2001whyQ} in 2001. In the subsequent years,
numerous other reconstructions of the finite-dimensional axioms followed
(e.g.~\cite{DAriano2007,Goyal2008,BD2010,GKS2010,CPDA2011,MMAP2013,Goyal2014,MM2016,Hoehn2017,SSC2018}),
which approached the reconstructive challenge using a wide variety
of mathematical and conceptual tools.

One of the desired characteristics of a reconstruction is that it
enables one to clearly grasp why the complex number system is appropriate
for the description of quantum phenomena. In this regard, reconstructions
of the Feynman formulation of quantum theory \cite{Feynman1948,FeH1965}
-- as opposed to the standard finite-dimensional Dirac--von Neumann
axioms -- are of special interest. The standard formalism attributes
quantum states to physical systems, stipulates how these evolve and
update in response to interactions with, and measurements upon, the
system. In contrast, the Feynman formulation eschews the notion of
quantum state and instead ascribes complex-valued amplitudes to \emph{paths}
that a system may take between detection events. The resulting formalism
is dramatically simpler than the standard state-based formalism. Its
three core rules prescribe how to combine these amplitudes according
to whether these paths are in series (1st rule) or in parallel (2nd),
and how to compute transition probabilities from these amplitudes
(3rd). Additional rules show, for example, how to combine amplitudes
belonging to transitions of non-interacting systems (which corresponds
to the usual tensor product rule in the Dirac--von Neumann axioms),
and how to handle identical particle systems. 

As shown in \cite{Tiko1988,Caticha1998,Tiko1999,GKS2010}, Feynman's
core rules possess striking symmetries (such as associativity, commutativity,
distributivity, and a product constraint for probabilities). Furthermore,
as shown in \cite{Caticha1998,GKS2010}, these symmetries can be justified
largely on \emph{operational} grounds, without explicit reference
to specific quantum phenomena, and moreover can be harnessed to reconstruct
Feynman's rules. However, these reconstructions leave something to
be desired. For example, at the outset, \cite{GKS2010} posits that
each path is represented by a pair of real numbers, thereby ruling
out algebraic possibilities beyond the complex numbers a priori. It
also employs a rather \emph{ad hoc} approach to the solution of the
symmetry equations, which leads to additional hitherto unresolved
issues \cite{Koepl2023gks}.

Our goal in the current paper is to harness the above-mentioned operationally-established
symmetries by determining a direct connection between these symmetries
and known foundational theorems in mathematics. There is rich historical
context in linear algebra, from associative division algebras (the
reals, complexes, and quaternions \cite{Frob1878}), multiplicative
sums of squares (also permitting nonassociative octonions \cite{Hurwitz1898}),
to a complete classification of algebras permitting a quadratic form
$p$ with product composition, $p\left(ab\right)=p\left(a\right)p\left(b\right)$
(composition algebras; \cite{Jacobson1958,Jacobson1981}). Our approach
rests on the observation that there is substantial overlap between
the symmetries and structures formulated in \cite{GKS2010,Goyal2014}
and the underlying axioms of the theorems that characterize composition
algebras. By judiciously expanding the set of operations formulated
in \cite{GKS2010}, and placing emphasis on specific types of paths,
we are able to show that complex numbers together with a small number
of other possible algebras are suitable to represent the operational
model\footnote{This in turn promises advancements for an existing body of related
research \cite{BCL1990,Goyal2012,Goyal2014,Goyal2015}, and may provide
a clearer understanding of alternative postulates introduced in \cite{GK2011,SK2019},
some of which faced critiques for their perceived unnecessary strength
(e.g.~\cite{HSP2014}).}.

The remainder of the paper is organized as follows. In section \ref{sec:Model},
we recap the original model from \cite{GKS2010} and develop select
algebraic properties. We avoid complete classification of this model
algebra, to not attempt to optimize what is supposed to be a building
block towards an unknown future completion. The question at hand is
then specified in section \ref{sec:Probability}, asking for probabilities
of possible outcomes given an experimental setup. To answer this question,
we introduce the choice representation of amplitudes in section \ref{sec:Amplitude-algebra},
supported by a simplicity argument towards answering the question
posed. Finally, in section \ref{sec:Born-rule}, we demonstrate that
allowable probabilities are quadratic in amplitudes, similar to the
Born rule from canonical physics, and classify allowable amplitude
algebras as the associative composition algebras over the reals. We
close with an outlook (section \ref{sec:Summary-and-outlook}) that
traces what was accomplished, points out nontrivial consequences of
seemingly minute tweaks to the model, and advertises exploration of
one specific generalization. Algebras that become permissible under
this abstract generalization match those from contemporary algebraic
representation of observed symmetries and structure of the fundamental
particles and forces in nature \cite{Furey2023rmap1,FuH2024}. This
is surprising -- to say the least -- given the different provenance.
Rather than taking quantum theory as a given, we invite to look at
it from a new angle altogether.

\subsection{Methodology}

The methodology in this paper is structured along three primary axes:
axiomatic definition, choice postulates, and derivation validation:
\begin{itemize}
\item \emph{Axioms} provide the foundational assumptions that determine
the model's relevance and scope.
\item \emph{Choices} are further postulates based on specific questions
asked from the model. They provide the template into which the inherent
properties of the phenomenon under study are to express themselves.
\item \emph{Derivations} then expose true statements that follow from axioms
and choices through deductive reasoning. This provides the basis for
evaluating choices for their logical coherence, to support their claimed
generality, and to ensure they do not conflict with the model.
\end{itemize}
Validation of axioms and choices ultimately resides in experimental
observation: Good choices yield useful answers, given the questions
asked. While mathematically, the difference between axioms and choices
may be little to none, the separation here aims at usefulness in physics:
\begin{itemize}
\item From the ground up: The axiomatic model is advertised as a primitive
building block in a way that promises broad utility towards a future
complete reconstruction of Quantum Mechanics. Clarity in axioms and
subsequent choices aims at providing researchers with an easier selection
on what to carry forward, what to modify, and what consequences may
follow.
\item With hindsight from canonical physics: After postulating a choice
representation of the model in terms of amplitudes, we will arrive
at a finite set of allowable amplitude algebras, which includes the
complex numbers. Any faithful reconstruction of canonical Quantum
Mechanics must explain the utility of the complexes. By highlighting
the choices we are making in representing our model algebra, tailored
to answer a specific question, we aim to clarify how to arrive at
the complexes.
\end{itemize}

\section{Model\protect\label{sec:Model}}

We adopt the operational model from \cite{GKS2010}, which constructs
experiments on quantum systems by arranging several measurements in
sequence. Each measurement comprises detectors, which are distinct
and non-filtering: Exactly one detector fires in each measurement,
corresponding to a measurement result. A sequence of measurement results
is then called a path. Sequential composition and simultaneous merging
of paths yield a model algebra with basic operations chaining and
coarsening, respectively. In this section, we develop the properties
and relations of this model algebra\footnote{The term ``algebra'' here is understood in the sense of universal
algebra as a space where certain operations act on some set of elements.
It is the ``logic'' of the model per \cite{GKS2010}.}, including inversion, reversal, and insertion.

Interactions between different quantum systems are assumed to be happening
between measurements. Together with spacetime and dynamics, interactions
are out of scope here.

\subsection{Measurements with possible measurement outcomes}

We denote an experiment as a sequence $\mathcal{M}$ with $\ell\geq2$
measurements $\mathbf{M}_{j}$, written
\begin{equation}
\mathcal{M}=\left[\mathbf{M}_{1},\mathbf{M}_{2},\ldots,\mathbf{M}_{\ell}\right].
\end{equation}
The source $\mathbf{S}$ is the first measurement in a sequence, $\mathbf{S}\left(\mathcal{M}\right)=\mathbf{M}_{1}$,
and the target $\mathbf{T}$ the last, $\mathbf{T}\left(\mathcal{M}\right)=\mathbf{M}_{\ell}$.
Each measurement comprises a finite number of detectors, $\mathbf{M}_{j}=\left\{ M_{j}^{1},M_{j}^{2},M_{j}^{3},\ldots\right\} $,
which form the set of possible outcomes for that measurement.

There are two classes of detectors $M_{j}^{r}$, atomic and non-atomic.
If all detectors in a measurement are atomic, we call the measurement
itself atomic. By postulate, we require the source and target of a
sequence to be atomic.

Two separate measurements are equal, $\mathbf{M}_{j}=\mathbf{M}_{k}$,
if they have identical possible outcomes and with that, are modeled
with the same set of detectors. A sequence that consists only of equal
measurements is called trivial.

We write a measurement's underlying atomic outcome elements as $m,m^{\prime},m^{\prime\prime},\ldots$.
Actual outcomes then are from a partition over these underlying elements.
For example, atomic outcomes are sets with one element, $\left\{ m\right\} $,
$\left\{ m^{\prime}\right\} $, $\left\{ m^{\prime\prime}\right\} $,
\ldots , and coarse-grained outcomes are sets with two or more elements,
$\left\{ m,m^{\prime}\right\} $, $\left\{ m,m^{\prime\prime}\right\} $,
$\left\{ m,m^{\prime},m^{\prime\prime}\right\} $, \ldots{}

We use raised-prefix notation $^{\alpha}\mathbf{M},{}^{\beta}\mathbf{M},{}^{\gamma}\mathbf{M},\ldots,{}^{\cup}\mathbf{M}$
to label different partitions over the same underlying element set
and call these measurements \emph{weakly equivalent}, $^{\alpha}\mathbf{M}\sim{}^{\beta}\mathbf{M}\sim{}^{\gamma}\mathbf{M}\sim\ldots\sim{}^{\cup}\mathbf{M}$.
Prefix $\alpha$ is reserved for atomic measurements, and prefix $\cup$
for measurements that are fully coarse-grained, i.e., consisting of
a single detector with an outcome that is the set of all elements.

For example, five possible measurements can be built over the set
with three underlying atomic outcome elements $m,m^{\prime},m^{\prime\prime}$,
corresponding to the five possible partitions of a set with three
elements. These measurements are weakly equivalent (see Figure \ref{fig:examplePartitionsOverThreeElements}):
\begin{figure}
\begin{centering}
\includegraphics[viewport=70bp 200bp 1050bp 742bp,clip,width=15cm]{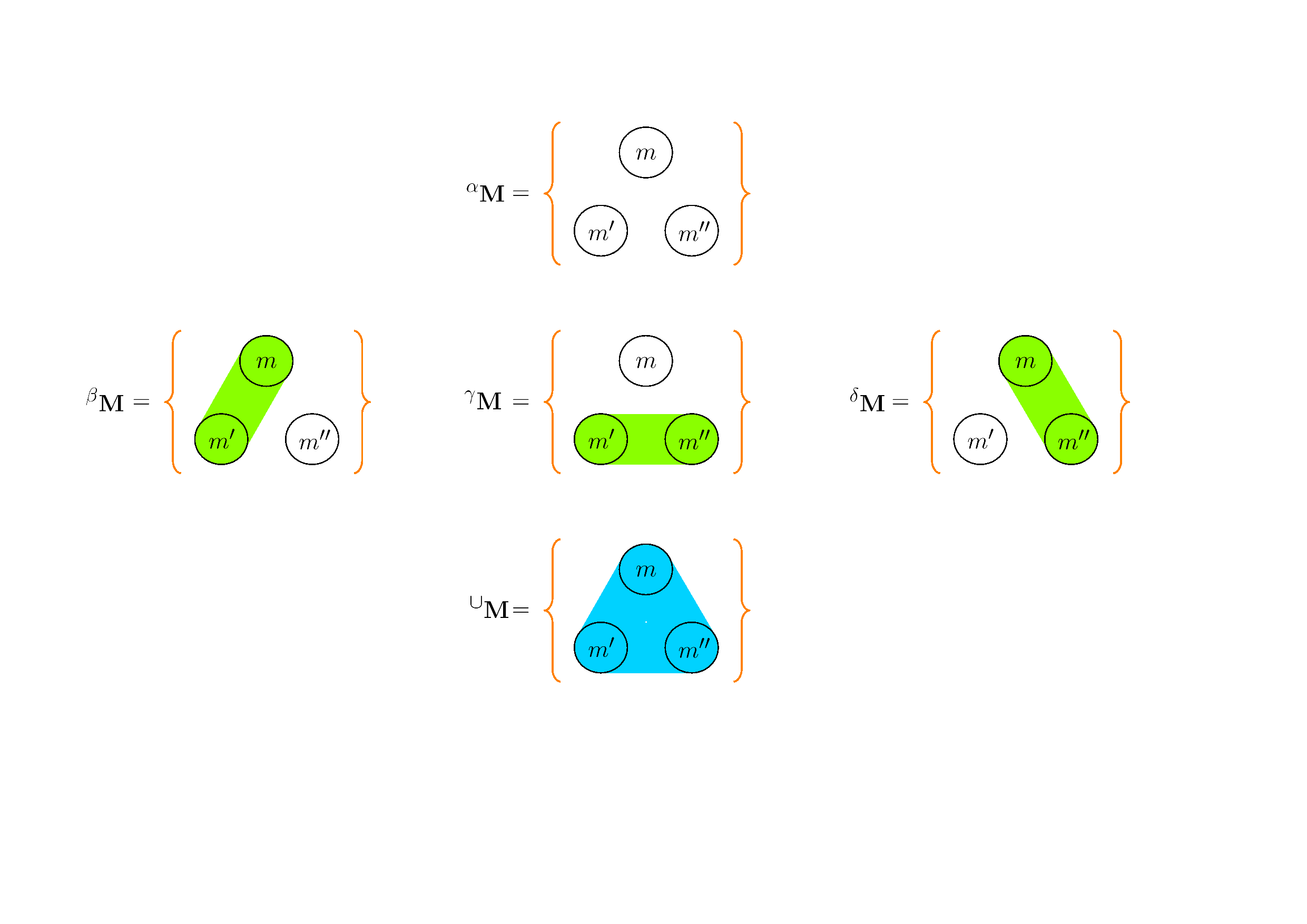}
\par\end{centering}
\caption{\protect\label{fig:examplePartitionsOverThreeElements}Given three
underlying atomic outcome elements $m,m^{\prime},m^{\prime\prime}$,
the five possible partitions are the measurements $^{\alpha}\mathbf{M},{}^{\beta}\mathbf{M},{}^{\gamma}\mathbf{M},{}^{\delta}\mathbf{M},{}^{\cup}\mathbf{M}$.
The elements in each partition are its detectors.}
\end{figure}
\begin{align}
^{\alpha}\mathbf{M} & =\left\{ \left\{ m\right\} ,\left\{ m^{\prime}\right\} ,\left\{ m^{\prime\prime}\right\} \right\} \nonumber \\
^{\beta}\mathbf{M} & =\left\{ \left\{ m,m^{\prime}\right\} ,\left\{ m^{\prime\prime}\right\} \right\} \nonumber \\
^{\gamma}\mathbf{M} & =\left\{ \left\{ m\right\} ,\left\{ m^{\prime},m^{\prime\prime}\right\} \right\} \label{eq:exampleDetectorsPartitionsOverThree}\\
^{\delta}\mathbf{M} & =\left\{ \left\{ m,m^{\prime\prime}\right\} ,\left\{ m^{\prime}\right\} \right\} \nonumber \\
^{\cup}\mathbf{M} & =\left\{ \left\{ m,m^{\prime},m^{\prime\prime}\right\} \right\} \nonumber 
\end{align}
\begin{align}
^{\alpha}\mathbf{M} & \sim{}^{\beta}\mathbf{M}\sim{}^{\gamma}\mathbf{M}\sim{}^{\delta}\mathbf{M}\sim{}^{\cup}\mathbf{M}.
\end{align}

\subsection{Paths with actual measurement results}

Given a measurement sequence $\mathcal{M}=\left[\mathbf{M}_{1},\mathbf{M}_{2},\ldots,\mathbf{M}_{\ell}\right]$,
paths $A,B,C,\ldots$ are sequences of measurement results from the
possible detectors at each measurement. The set of all allowable paths
over $\mathcal{M}$ is written as $\mathcal{S_{M}}$.

Given a path $A\in\mathcal{S_{M}}$, we write $A_{j}$ for the detector
result at measurement $\mathbf{M}_{j}$, i.e., $A_{j}\in\mathbf{M}_{j}$,
\begin{equation}
A=\left[A_{1},A_{2},\ldots,A_{\ell}\right].
\end{equation}
Just as the source and target in a path are the respective measurements
in a sequence, $\mathbf{S}\left(A\right)=\mathbf{S}\left(\mathcal{M}\right)$
and $\mathbf{T}\left(A\right)=\mathbf{T}\left(\mathcal{M}\right)$,
we call the first and last measurement outcome in a path the source
result $S\left(A\right)=A_{1}$ and target result $T\left(A\right)=A_{\ell}$.
Because the source and target of a measurement sequence must be atomic,
this will be the case for source and target results.

\subsubsection{Preparation and selection}

An experiment always consists of at least two measurements ($\ell\geq2$).
The first measurement, $\mathbf{M}_{1}$, is required to be atomic
and prepares the system such that the experiment is well-defined and
can be repeated. The last measurement, $\mathbf{M}_{\ell}$, is also
atomic and yields a specific outcome. Therefore, measurement sequences
$\left[\mathbf{M}_{1},\mathbf{M}_{2}\right]$ of length $\ell=2$,
with paths $\left[M_{1},M_{2}\right]$, are the shortest possible
sequences.

\subsubsection{Cyclic, symmetric (palindromic), and trivial paths}

A path $^{\circ}\negthinspace A$ is cyclic if its source and target
are identical, i.e., $\mathbf{M}_{1}=\mathbf{M}_{\ell}$ with $S\left(^{\circ}\negthinspace A\right)=T\left(^{\circ}\negthinspace A\right)$.
Furthermore, cyclic paths are symmetric (or palindromic), $^{\circledcirc}\negthinspace A$,
if their measurements and outcomes are identical in the opposite order:
$\mathbf{M}_{1}=\mathbf{M}_{\ell}$ with result $^{\circledcirc}\negthinspace A_{1}=^{\circledcirc}\negthinspace A_{\ell}$,
$\mathbf{M}_{2}=\mathbf{M}_{\ell-1}$ with $^{\circledcirc}\negthinspace A_{2}=^{\circledcirc}\negthinspace A_{\ell-1}$,
and so on. Note that the definitions for cyclic and symmetric paths
require identical outcomes \emph{and also} equal measurements at the
relevant sequence positions: Paths over cyclic (or symmetric) measurement
sequences generally do not have cyclic (or symmetric) outcomes, and
conversely, a path with measurement outcomes matching these patterns
does not require matching measurement patterns due to the weak equivalence
freedom for measurements.

Take, for example, the two measurements $^{\alpha}\mathbf{M}=\left\{ \left\{ m\right\} ,\left\{ m^{\prime}\right\} ,\left\{ m^{\prime\prime}\right\} \right\} $
and $^{\delta}\mathbf{M}=\left\{ \left\{ m,m^{\prime\prime}\right\} ,\left\{ m^{\prime}\right\} \right\} $
from equation (\ref{eq:exampleDetectorsPartitionsOverThree}) above.
Both have detector $M^{\prime}=\left\{ m^{\prime}\right\} $ as a
possible outcome. The $^{\alpha}\mathbf{M}$ and $^{\delta}\mathbf{M}$
are weakly equivalent but not equal. Now take two more arbitrary measurements,
$\mathbf{N}$ and $\mathbf{O}$, with results $N\in\mathbf{N}$ and
$O\in\mathbf{O}$, respectively, and form a measurement sequence $\left[\mathbf{N},{}^{\alpha}\mathbf{M},\mathbf{O},{}^{\delta}\mathbf{M},\mathbf{N}\right]$.
This sequence allows for results $\left[N,M^{\prime},O,M^{\prime},N\right]$,
which would be considered cyclic but not symmetric because $^{\alpha}\mathbf{M}\neq{}^{\delta}\mathbf{M}$.

Finally, a trivial path consists only of identical measurement results
over equal measurements.

\subsection{Model algebra}

\subsubsection{\protect\label{subsec:Chaining-and-coarsening}Chaining and coarsening}

We follow the original approach in \cite{GKS2010} and develop algebraic
relations for measurement sequences and paths. In order to avoid confusion
with established mathematical terms like ``sequence algebra'' or
``path algebra,'' we adopt a generic term \emph{model algebra} here.
It simply denotes algebraic relations that trace from the operational
model in this paper without attempting to specify their general context\footnote{That is not to say that exploring the model algebra by itself in the
context of established mathematical frameworks would not be interesting.
Doing so is bound to provide fruitful insight. As we make certain
choices in this paper, we aim to allow for the widest possible interpretation
and generalization of our results. Narrowing the model algebra at
this point is unneeded and, therefore, unwanted for our purpose.}.

The space of all measurements is now written as $\mathbb{M}$. Experiments
then are from the possible sequences of measurements, written $\left[\mathbb{M}_{j}\right]_{j\geq2}$,
or simply $\left[\mathbb{M}\right]$. Two measurement sequences $\mathcal{M}^{\mathrm{A}}=\left[\mathbf{M}_{1}^{\mathrm{A}},\mathbf{M}_{2}^{\mathrm{A}},\ldots,\mathbf{M}_{\ell_{\mathrm{A}}}^{\mathrm{A}}\right]$
and $\mathcal{M}^{\mathrm{B}}=\left[\mathbf{M}_{1}^{\mathrm{B}},\mathbf{M}_{2}^{\mathrm{B}},\ldots,\mathbf{M}_{\ell_{\mathrm{B}}}^{\mathrm{B}}\right]$,
$\mathcal{M}^{\mathrm{A}},\mathcal{M}^{\mathrm{B}}\in\left[\mathbb{M}\right]$,
where the target of $\mathcal{M}^{\mathrm{A}}$ is equal to the source
of $\mathcal{M}^{\mathrm{B}}$, may be composed sequentially to form
a new sequence $\mathcal{M}^{\mathrm{C}}$:
\begin{align}
\cdot & :\left[\mathbb{M}\right]\times\left[\mathbb{M}\right]\rightharpoonup\left[\mathbb{M}\right],\nonumber \\
\mathcal{M}^{\mathrm{A}},\mathcal{M}^{\mathrm{B}},\mathcal{M}^{\mathrm{C}} & \in\left[\mathbb{M}\right],\\
 & \textrm{with }\mathbf{T}\left(\mathcal{M^{\mathrm{A}}}\right)=\mathbf{S}\left(\mathcal{M}^{\mathrm{B}}\right),\nonumber \\
\mathcal{M}^{\mathrm{C}}=\mathcal{M}^{\mathrm{A}}\cdot\mathcal{M}^{\mathrm{B}} & :=\left[\mathbf{M}_{1}^{\mathrm{A}},\ldots,\mathbf{M}_{\ell_{\mathrm{A}}}^{\mathrm{A}}=\mathbf{M}_{1}^{\mathrm{B}},\ldots,\mathbf{M}_{\ell_{\mathrm{B}}}^{\mathrm{B}}\right],\nonumber \\
\ell_{\mathrm{C}} & =\ell_{\mathrm{A}}+\ell_{\mathrm{B}}-1.
\end{align}
This sequential composition operation is called \emph{chaining}. Because
the operation is only allowed when the target of the first operand
is equal to the source of the second, $\mathbf{M}_{\ell_{\mathrm{A}}}^{\mathrm{A}}=\mathbf{M}_{1}^{\mathrm{B}}$,
it is a partial binary operation ($\rightharpoonup$) on the space
of all measurement sequences $\left[\mathbb{M}\right]$.

If paths $A,B$ exist over these measurement sequences, $A\in\mathcal{S}_{\mathcal{M}^{\mathrm{A}}}$,
$B\in\mathcal{S}_{\mathcal{M}^{\mathrm{B}}}$, where likewise the
target of $A$ matches the source of $B$, then the same operation
is defined for measurement paths as well:
\begin{align}
\cdot & :\mathcal{S}_{\left[\mathbb{M}\right]}\times\mathcal{S}_{\left[\mathbb{M}\right]}\rightharpoonup\mathcal{S}_{\left[\mathbb{M}\right]},\nonumber \\
A & \in\mathcal{S}_{\mathcal{M}^{\mathrm{A}}},\nonumber \\
B & \in\mathcal{S}_{\mathcal{M}^{\mathrm{B}}},\\
 & \begin{array}{rl}
\textrm{with} & \mathbf{T}\left(\mathcal{M^{\mathrm{A}}}\right)=\mathbf{S}\left(\mathcal{M}^{\mathrm{B}}\right),\\
 & \mathbf{t}\left(A\right)=\mathbf{s}\left(B\right),
\end{array}\nonumber \\
A\cdot B & =\left[A_{1},A_{2},\ldots,A_{\ell_{A}}=B_{1},B_{2},\ldots,B_{\ell_{B}}\right],\nonumber \\
\ell_{A\cdot B} & =\ell_{A}+\ell_{B}-1.
\end{align}
\begin{figure}
\begin{centering}
\includegraphics[viewport=50bp 180bp 1141bp 560bp,clip,width=15cm]{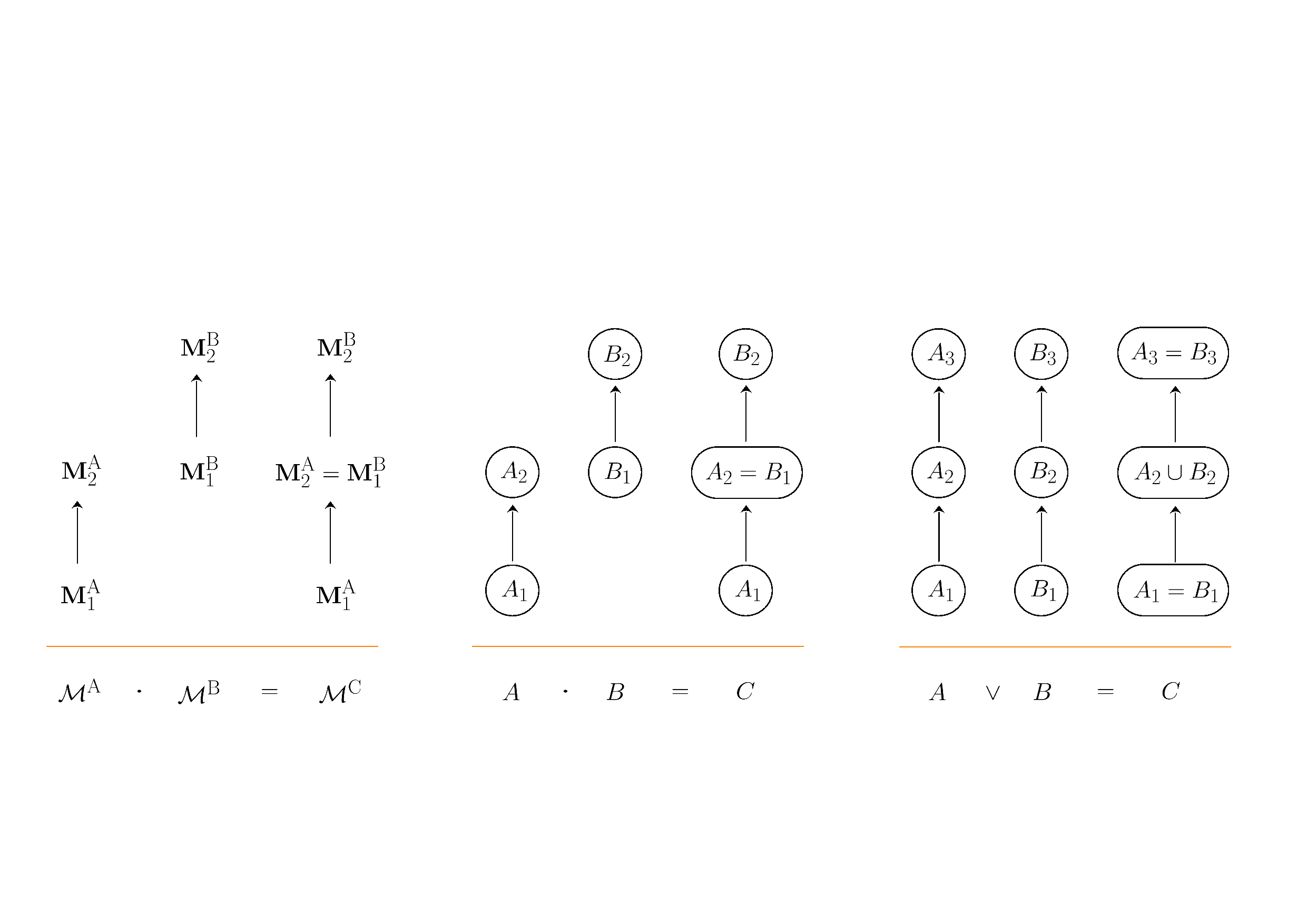}
\par\end{centering}
\caption{\protect\label{fig:chainingAndCoarsening}Chaining ($\cdot$) and
coarsening ($\vee$) of sequences of measurements (experiments; $\mathcal{M}^{\mathrm{A}}\cdot\mathcal{M}^{\mathrm{B}}$)
and measurement results (paths; $A\cdot B$, $A\vee B$).}
\end{figure}

Another operation on paths is \emph{coarsening}, which is a form of
simultaneous merging of possible detector outcomes and measurement
results on a path. Given a measurement sequence $\mathcal{M}=\left[\mathbf{M}_{1},\mathbf{M}_{2},\ldots,\mathbf{M}_{\ell}\right]$,
coarsening modifies exactly one of these $\mathbf{M}_{j}\in\mathcal{M}$
that is neither source or target\footnote{Source and target must be atomic and, therefore, could not result
in an allowable path if either were to be coarsened.}, $1<j<\ell$, by replacing it with a coarsened version, $^{\vee}\mathbf{M}_{j}\in\mathbb{M}$.
All detector outcomes of $\mathbf{M}_{j}$ are subsets of the outcomes
of the newly formed, different measurement $^{\vee}\mathbf{M}_{j}$
of lesser cardinality, $\left|^{\vee}\mathbf{M}_{j}\right|<\left|\mathbf{M}_{j}\right|$,
and the $^{\vee}\mathbf{M}_{j}$ are always non-atomic. This requires
$\mathbf{M}_{j}$ and $^{\vee}\mathbf{M}_{j}$ to be weakly equivalent
but not equal:
\begin{equation}
\mathbf{M}_{j}\sim{}^{\vee}\mathbf{M}_{j},\,\,\,\,\,\,\,\,\mathbf{M}_{j}\neq{}^{\vee}\mathbf{M}_{j},\,\,\,\,\,\,\,\,\left|^{\vee}\mathbf{M}_{j}\right|<\left|\mathbf{M}_{j}\right|.
\end{equation}
We write $\mathcal{M}^{\vee\left\{ j\right\} }$ for the sequence
that differs from $\mathcal{M}$ exactly by replacing the measurement
$\mathbf{M}_{j}$ with $^{\vee}\mathbf{M}_{j}$.

Formally, take two paths $A,B\in\mathcal{S}_{\mathcal{M}}$ that differ
only at $\mathbf{M}_{j}$ by $A_{j},B_{j}\in\mathbf{M}_{j}$, $A_{j}\in A$,
$B_{j}\in B$, $A_{j}\cap B_{j}=\emptyset$, and form the new measurement
$^{\vee}\mathbf{M}_{j}$ by coarsening these two detector outcomes
to become a single new coarse-grained outcome $A_{j}\cup B_{j}\in{}^{\vee}\mathbf{M}_{j}$.
Coarsening then yields a path that contains this coarsened outcome
as a result:
\begin{align}
\vee & :\mathcal{S}_{\left[\mathbb{M}\right]}\times\mathcal{S}_{\left[\mathbb{M}\right]}\rightharpoonup\mathcal{S}_{\left[\mathbb{M}\right]},\nonumber \\
A,B & \in\mathcal{S}_{\mathcal{M}},\nonumber \\
C & \in\mathcal{S}_{\mathcal{M}^{\lor\left\{ j\right\} }},\\
 & \begin{array}{rl}
\textrm{with} & 1<j<\ell_{A}=\ell_{B},\\
 & A_{j},B_{j}\in\mathbf{M}_{j}\in\mathcal{M},\\
 & A_{j}\cap B_{j}=\emptyset,\\
 & C_{k}:=A_{k}=B_{k}\,\left(\forall k\neq j\right),\\
 & C_{j}:=A_{j}\cup B_{j}\in{}^{\vee}\mathbf{M}_{j}\in\mathcal{M}^{\lor\left\{ j\right\} },\\
 & \left|^{\vee}\mathbf{M}_{j}\right|=\left|\mathbf{M}_{j}\right|-1,
\end{array}\nonumber \\
A\vee B & :=C.\nonumber 
\end{align}
Similar to chaining, coarsening $\vee:\mathcal{S}_{\mathbb{\left[\mathbb{M}\right]}}\times\mathcal{S}_{\mathbb{\left[\mathbb{M}\right]}}\rightharpoonup\mathcal{S}_{\left[\mathbb{M}\right]}$
is a partial binary operation, i.e., it is only defined on a subset
of compatible elements from $\mathcal{S}_{\left[\mathbb{M}\right]}$.
See Figure \ref{fig:chainingAndCoarsening} for an illustration.

\subsubsection{Chaining is associative but partially noncommutative}

Sequences of measurements and outcomes are strictly ordered. The same
experiment can be conceptually divided in different ways. When chaining
three or more sequences, which sequence pairs are chained first is
irrelevant: The operation is associative. For suitable $\mathcal{M}^{\mathrm{A}},\mathcal{M}^{\mathrm{B}},\mathcal{M}^{\mathrm{C}}\in\mathbb{\left[\mathbb{M}\right]}$,
$A\in\mathcal{S}_{\mathcal{M}^{\mathrm{A}}}$, $B\in\mathcal{S}_{\mathcal{M}^{\mathrm{B}}}$,
$C\in\mathcal{S}_{\mathcal{M}^{\mathrm{C}}}$, there is
\begin{align}
\left(\mathcal{M}^{\mathrm{A}}\cdot\mathcal{M}^{\mathrm{B}}\right)\cdot\mathcal{M}^{\mathrm{C}} & =\mathcal{M}^{\mathrm{A}}\cdot\left(\mathcal{M}^{\mathrm{B}}\cdot\mathcal{M}^{\mathrm{C}}\right),\\
\left(A\cdot B\right)\cdot C & =A\cdot\left(B\cdot C\right).
\end{align}

However, chaining is generally not commutative. This is trivially
the case when only one of the commuted expressions for measurement
sequences $\left\{ \mathcal{M}^{\mathrm{A}}\cdot\mathcal{M}^{\mathrm{B}},\mathcal{M}^{\mathrm{B}}\cdot\mathcal{M}^{\mathrm{A}}\right\} $
or paths $\left\{ A\cdot B,B\cdot A\right\} $ is defined: Only because
the target of a $\mathcal{M}^{\mathrm{A}}$ matches the source of
$\mathcal{M}^{\mathrm{B}}$, permitting chaining $\mathcal{M}^{\mathrm{A}}\cdot\mathcal{M}^{\mathrm{B}}$,
this does not have to be true when commuting the two, permitting $\mathcal{M}^{\mathrm{B}}\overset{?}{\cdot}\mathcal{M}^{\mathrm{A}}$
in principle.

In the other case, when both commuted expressions $\mathcal{M}^{\mathrm{A}}\cdot\mathcal{M}^{\mathrm{B}}$
and $\mathcal{M}^{\mathrm{B}}\cdot\mathcal{M}^{\mathrm{A}}$ are defined,
there is still no a priori basis for this symmetry. We don't need
to assume that changing the order of measurements or changing the
order of measurement outcomes never affects the observation.

Since not every commuted expression is allowable, we call this property
of chaining \emph{partial noncommutativity}: Where defined, chaining
may or may not be commutative.

\subsubsection{Coarsening is commutative and partially associative}

Coarsening acts on the set of possible measurement outcomes by merging
two detectors. When a measurement is performed, the corresponding
merged results in a path are yielded. Since there is no ordering of
detectors that make up a measurement, the coarsening operation is
trivially commutative: For suitable $\mathcal{M}\in\mathbb{\left[\mathbb{M}\right]}$,
$A,B,C\in\mathcal{S}_{\mathcal{M}}$, there is
\begin{equation}
A\vee B=B\vee A.
\end{equation}
Where allowable, coarsening is also associative:
\begin{equation}
\left(A\vee B\right)\vee C=A\vee\left(B\vee C\right).
\end{equation}
To account that either side of this associativity relation may be
undefined, we call this property \emph{partial associativity}: Where
defined, coarsening is always associative. For both sides of the associativity
relation to be defined, the three paths $A,B,C$ must differ exactly
(and only) at the same sequence index $j$, i.e., $\left(A\vee B\right)\in\mathcal{S}_{\mathcal{M}^{\lor\left\{ j\right\} }}$
and $\left(B\vee C\right)\in\mathcal{S}_{\mathcal{M}^{\lor\left\{ j\right\} }}$.

\subsubsection{Chaining distributes over coarsening}

\begin{figure}
\begin{centering}
\includegraphics[viewport=130bp 170bp 1100bp 670bp,clip,width=15cm]{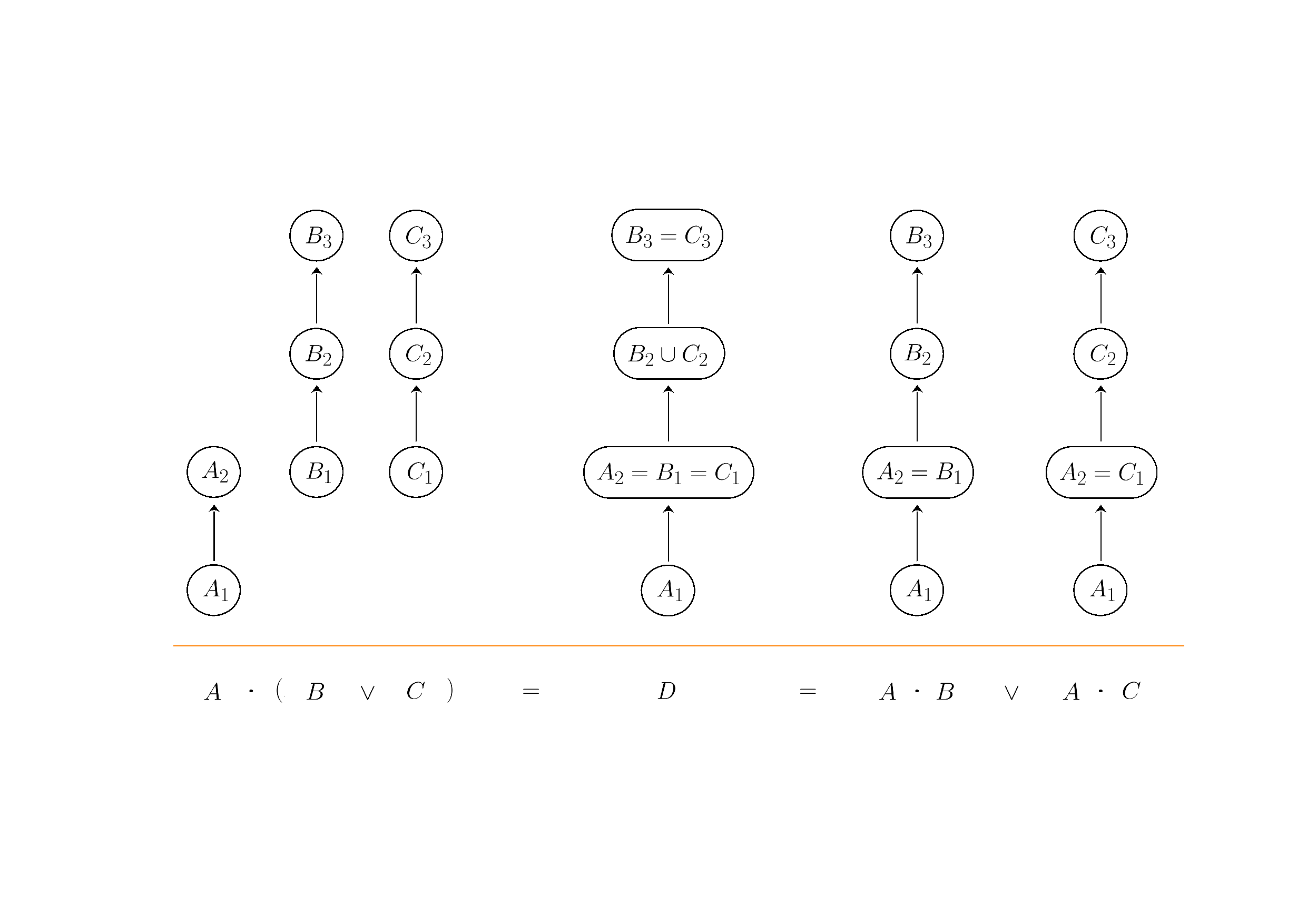}
\par\end{centering}
\caption{\protect\label{fig:chainingDistributesOverCoarsening}Chaining distributes
over coarsening}
\end{figure}
It can be readily seen that chaining is distributive over coarsening
(figure \ref{fig:chainingDistributesOverCoarsening}): For $\mathcal{M}^{\mathrm{A}},\mathcal{M}^{\mathrm{BC}}\in\mathbb{\left[\mathbb{M}\right]}$,
$A\in\mathcal{S}_{\mathcal{M}^{\mathrm{A}}}$, $B,C\in\mathcal{S}_{\mathcal{M}^{\mathrm{BC}}}$,
there is
\begin{equation}
A\cdot\left(B\vee C\right)=A\cdot B\vee A\cdot C.
\end{equation}
The left-hand side of the left-distributivity relation chains a path
over a measurement sequence $\mathcal{M}^{\mathrm{A}}$ to a path
over $\mathcal{M}^{\mathrm{BC}\lor\left\{ j\right\} }$, i.e., over
a sequence of measurements that differs from the original $\mathcal{M}^{\mathrm{BC}}$
only at an index $j\neq1,\ell$. With that, the source of $B,C$ is
the same as that of $B\vee C$. Conversely, the right-hand side of
the left-distributivity relation extends $B$ and $C$ by an identical
path $A$ that is not subject to coarsening.

A similar argument applies to right-distributivity:
\begin{equation}
\left(B\vee C\right)\cdot A=B\cdot A\vee C\cdot A.
\end{equation}

\subsubsection{Reversal}

Reversal takes a sequence of measurements $\mathcal{M}=\left[\mathbf{M}_{1},\mathbf{M}_{2},\ldots,\mathbf{M}_{\ell}\right]\in\mathbb{\left[\mathbb{M}\right]}$
or measurement results $A=\left[A_{1},A_{2},\ldots,A_{\ell}\right]\in\mathcal{S}_{\mathcal{M}}$
and puts all elements in the opposite order:
\begin{align}
\overline{\phantom{{A}}} & :\mathbb{\left[\mathbb{M}\right]}\rightarrow\mathbb{\left[\mathbb{M}\right]},\nonumber \\
\overline{\mathcal{M}} & =\left[\mathbf{M}_{\ell},\mathbf{M}_{\ell-1},\ldots,\mathbf{M}_{1}\right],
\end{align}
\begin{align}
\overline{\phantom{{A}}} & :\mathcal{S}_{\mathcal{M}}\rightarrow\mathcal{S}_{\overline{\mathcal{M}}},\nonumber \\
\overline{A} & =\left[A_{\ell},A_{\ell-1},\ldots,A_{1}\right].
\end{align}
This operation replaces the source and target of the sequence, $\mathbf{T}\left(\overline{\mathcal{M}}\right)=\mathbf{S}\left(\mathcal{M}\right)$,
$\mathbf{S}\left(\overline{\mathcal{M}}\right)=\mathbf{T}\left(\mathcal{M}\right)$,
$T\left(\overline{A}\right)=S\left(A\right)$, and $S\left(\overline{A}\right)=T\left(A\right)$.
Reversal is an involution,
\begin{equation}
\overline{\overline{\mathcal{M}}}=\mathcal{M},
\end{equation}
that distributes over coarsening. It reverses the order of arguments
when distributing over chaining (sometimes called ``anti-involution''
or ``anti-isomorphism''),
\begin{align}
\overline{A\vee B} & =\overline{A}\vee\overline{B},\\
\overline{\mathcal{M}^{\mathrm{A}}\cdot\mathcal{M}^{\mathrm{B}}} & =\overline{\mathcal{M}^{\mathrm{B}}}\cdot\overline{\mathcal{M}^{\mathrm{A}}}.\\
\overline{A\cdot B} & =\overline{B}\cdot\overline{A}.
\end{align}
The reverse of a path is also called its adjoint.

Symmetric sequences are self-adjoint, i.e., they remain invariant
under reversal. For symmetric $^{\circledcirc}\negthinspace\mathcal{M}\in\mathbb{\left[\mathbb{M}\right]}$
and $^{\circledcirc}\negthinspace A\in\mathcal{S}_{^{\circledcirc}\negthinspace\mathcal{M}}$
there is:
\begin{align}
\overline{^{\circledcirc}\negthinspace\mathcal{M}} & \,=\,^{\circledcirc}\negthinspace\mathcal{M},\\
\overline{^{\circledcirc}\negthinspace A} & \,=\,^{\circledcirc}\negthinspace A.
\end{align}

\subsubsection{\protect\label{subsec:Inverse-operations-unchaining}Inverse operations
unchaining and refinement}

When we build a sequence of measurements individually, we have complete
knowledge of all steps in its construction. Just as we can form new
measurement sequences by chaining them together or coarsening the
set of some detector outcomes, we can recover the prior sequences
by simply not doing these operations. The same applies to paths once
their underlying prior measurement setup has been recovered. In the
model algebra, we reflect this by formally introducing inverse operations
unchaining and refinement.

After chaining measurement sequences, $\mathcal{M}^{\mathrm{A}}\cdot\mathcal{M}^{\mathrm{B}}=\mathcal{M}^{\mathrm{C}}$
or result paths $A\cdot B=C$, we can start with $\mathcal{M}^{\mathrm{C}}$
or $C$ and formally recover the prior (noncommutative) arguments
by left- and right-unchaining, $\setminus$ and $/$, respectively:
\begin{align}
/,\setminus & :\mathbb{\left[\mathbb{M}\right]}\times\mathbb{\left[\mathbb{M}\right]}\rightharpoonup\mathbb{\left[\mathbb{M}\right]},\nonumber \\
\mathcal{M}^{\mathrm{C}}/\mathcal{M}^{\mathrm{B}} & =\left(\mathcal{M}^{\mathrm{A}}\cdot\mathcal{M}^{\mathrm{B}}\right)/\mathcal{M}^{\mathrm{B}}=\mathcal{M}^{\mathrm{A}},\\
\mathcal{M}^{\mathrm{A}}\setminus\mathcal{M}^{\mathrm{C}} & =\mathcal{M}^{\mathrm{A}}\setminus\left(\mathcal{M}^{\mathrm{A}}\cdot\mathcal{M}^{\mathrm{B}}\right)=\mathcal{M}^{\mathrm{B}},
\end{align}
\begin{align}
/,\setminus & :\mathcal{S}_{\mathbb{\left[\mathbb{M}\right]}}\times\mathcal{S}_{\mathbb{\left[\mathbb{M}\right]}}\rightharpoonup\mathcal{S}_{\mathbb{\left[\mathbb{M}\right]}},\nonumber \\
C/B & =\left(A\cdot B\right)/B=A,\label{eq:rightUnchainingDef}\\
A\setminus C & =A\setminus\left(A\cdot B\right)=B.\label{eq:leftUnchainingDef}
\end{align}
Commuting or associating unchaining is generally undefined.

For coarsened paths $A\vee B=C$ (where $A,B\in\mathcal{S}_{\mathcal{M}}$,
$C\in\mathcal{S}_{\mathcal{M}^{\lor\left\{ j\right\} }}$), we start
with $C$ and recover the prior argument by refinement $\wedge$:
\begin{align}
\wedge & :\mathcal{S}_{\mathcal{M}^{\lor\left\{ j\right\} }}\times\mathcal{S}_{\mathcal{M}}\rightharpoonup\mathcal{S}_{\mathbb{\mathcal{M}}},\nonumber \\
C\wedge B & =A,\label{eq:refinementDef}\\
C\wedge A & =B.
\end{align}
Because coarsening does not depend on the order of its arguments,
we do not need to distinguish between a left- and right-inverse operation.
Just as for unchaining, commuting or associating refinement is generally
undefined.

These inverses are partial binary operations only on a subset of allowable
pairs of measurement sequences and result paths. We point out that
inverses must nevertheless be \emph{partially well-defined}, i.e.,
the operations are unique wherever they do exist\footnote{A conceptual example would be squaring in positive whole numbers $\mathbb{Z}^{+}$,
for which inverses are well-defined where they exist. In contrast,
the inverse operation to squaring in the integers $\mathbb{Z}$ would
not be well-defined in this sense: Where allowed, it generally permits
\emph{two} possible solutions (the only exception being 0).}: We have full knowledge of the details of a given sequence and know
which parts can be removed; and with that, we know what happens when
removing some of these parts.

\subsubsection{Insertion}

Insertion is now introduced as an operation that expands a given sequence
by adding measurements to a sequence or results to a path.

First, we define insertion in the most general case, adding a single
measurement or result at a specified sequence index. Given a measurement
sequence, $\mathcal{M}=\left[\mathbf{M}_{1},\mathbf{M}_{2},\ldots,\mathbf{M}_{\ell}\right]\in\left[\mathbb{M}\right]$,
we insert a new measurement $\mathbf{M}_{\lambda}\in\mathbb{M}$ at
position $j\in\left\{ 2\ldots\ell\right\} \in\mathbb{Z}^{+}$ by shifting
the existing measurement $\mathbf{M}_{j}\in M$ at $j$ up by one
index, and then adding the new measurement $\mathbf{M}_{\lambda}$
at the position occupied initially by $\mathbf{M}_{j}$:
\begin{align}
\downarrow & :\left[\mathbb{M}\right]\times\mathbb{M}\times\mathbb{Z}^{+}\rightarrow\left[\mathbb{M}\right],\nonumber \\
\mathcal{M}\downarrow_{j}\mathbf{M}_{\lambda} & :=\left[\mathbf{M}_{1},\ldots,\mathbf{M}_{j-1},\mathbf{M}_{\lambda},\mathbf{M}_{j},\ldots,\mathbf{M}_{\ell}\right].
\end{align}
The operation leaves the source and target of the original sequence
$\mathcal{M}$ unchanged but increases its length by 1,
\begin{equation}
\ell_{\mathcal{M}\downarrow_{j}\mathbf{M}^{\lambda}}=\ell_{\mathcal{M}}+1.
\end{equation}

Similarly, for measurement results $A=\left[A_{1},A_{2},\ldots,A_{\ell}\right]\in\mathcal{S}_{\mathcal{M}}$
we insert a new result $A_{\lambda}\in\mathbf{M}_{\lambda}$ into
the path by:
\begin{align}
\downarrow & :\mathcal{S}_{\mathcal{M}}\rightarrow\mathcal{S}_{\mathcal{M}\downarrow_{j}\mathbf{M}_{\lambda}},\nonumber \\
A\downarrow_{j}A_{\lambda} & :=\left[A_{1},\ldots,A_{j-1},A_{\lambda},A_{j},\ldots,A_{\ell}\right].
\end{align}
For the sake of readability, this compact notation ``$\downarrow\,:\mathcal{S}_{\mathcal{M}}\rightarrow\mathcal{S}_{\mathcal{M}\downarrow_{j}\mathbf{M}_{\lambda}}$''
accounts for inserting a new result $A_{\lambda}$ into the original
path $\mathcal{S}_{\mathcal{M}}$ at position $j$, \emph{as well
as} a measurement $\mathbf{M}_{\lambda}$ that turns $\mathcal{M}$
into $\mathcal{M}\downarrow_{j}\mathbf{M}_{\lambda}$.

When inserting sequences of measurements and results, we use a similar
notation as for the insertion of individual ones. Inserting a cyclic
measurement sequence $^{\circ}\negthinspace\mathcal{M}^{\Lambda}$
after the sequence $\mathcal{M}^{\mathrm{J}}$ of a sequence chain
$\mathcal{M}^{\mathrm{1}}\cdot\ldots\cdot\mathcal{M}^{\mathrm{L}}$
is written as:
\begin{align}
\downarrow & :\left[\mathbb{M}\right]\times\left[\mathbb{M}\right]\times\mathbb{Z}^{+}\rightarrow\left[\mathbb{M}\right],\nonumber \\
\mathcal{M}\downarrow_{\mathrm{J}}{}^{\circ}\negthinspace\mathcal{M}^{\Lambda} & =\mathcal{M}^{1}\cdot\mathcal{M}^{2}\cdot\ldots\cdot\mathcal{M}^{\mathrm{J}-1}\cdot{}^{\circ}\negthinspace\mathcal{M}^{\Lambda}\cdot\mathcal{M}^{\mathrm{J}}\cdot\ldots\cdot\mathcal{M}^{\mathrm{L}}.
\end{align}
Chaining requires $\mathbf{T}\left(\mathcal{M}^{\mathrm{J}-1}\right)=\mathbf{S}\left(^{\circ}\negthinspace\mathcal{M}^{\Lambda}\right)=\mathbf{T}\left(^{\circ}\negthinspace\mathcal{M}^{\Lambda}\right)=\mathbf{S}\left(\mathcal{M}^{\mathrm{J}}\right)$
and this operation is only allowed for cyclic $^{\circ}\negthinspace\mathcal{M}^{\Lambda}$.
It increases sequence length by all elements of $^{\circ}\negthinspace\mathcal{M}^{\Lambda}$
that are not its source (or target):
\begin{equation}
\ell_{\mathcal{M}\downarrow_{\mathrm{J}}{}^{\circ}\negthinspace\mathcal{M}^{\Lambda}}=\ell_{\mathcal{M}}+\ell_{^{\circ}\negthinspace\mathcal{M}^{\Lambda}}-1.
\end{equation}
Similarly for measurement results $A^{\Lambda}=\left[A_{1}^{\Lambda},A_{2}^{\Lambda},\ldots,A_{\ell_{\Lambda}}^{\Lambda}\right]\in\mathcal{S}_{^{\circ}\negthinspace\mathcal{M}^{\Lambda}}$:
\begin{align}
\downarrow & :\mathcal{S}_{\mathcal{M}}\rightarrow\mathcal{S}_{\mathcal{M}\downarrow_{\mathrm{J}}{}^{\circ}\negthinspace\mathcal{M}^{\Lambda}},\nonumber \\
A\downarrow_{\mathrm{J}}A^{\Lambda} & :=A^{1}\cdot\ldots A^{\mathrm{J}-1}\cdot A^{\Lambda}\cdot A^{\mathrm{J}}\ldots\cdot A^{\mathrm{L}}.
\end{align}

\section{\protect\label{sec:Probability}Probability}

So far, we have laid out an operational model with experiments that
are sequences of measurements, possible detector outcomes, and paths
that are sequences of measurement results when an experiment is performed.
We have traced algebraic properties from select operations on these
sequences: chaining, coarsening, their inverses, reversal, and insertions.
In this model algebra, the coarsening operation is motivated from
Quantum Mechanics: We want to permit detectors that fire on one of
multiple possible, distinct underlying atomic outcome elements, where
it does not need to be known which one exactly triggered the signal.

As a next step, we clarify the questions one may want to ask from
the operational model: In Quantum Mechanics, the experimenter asks
for probability distributions across possible measurement outcomes,
given some preparation. The concept of probability is extraneous to
our model; it is the experimenter who imposes it as mathematical construct
in order to formulate the question. Nature, in turn, yields an answer
given this constraint. If the experimenter is asking good questions,
and the model represents nature, then the answers will be useful.

\subsection{Paths have probability and establish closure}

A general path $A=\left[A_{1},A_{2},\ldots,A_{\ell}\right]\in\mathcal{S}_{\mathcal{M}}$
over measurements $\mathcal{M}=\left[\mathbf{M}_{1},\mathbf{M}_{2},\ldots,\mathbf{M}_{\ell}\right]\in\left[\mathbb{M}\right]$,
$\ell\geq2$, will now be associated with a probability $P\in\left[0,1\right]\in\mathbb{R}$.
Each measurement $\mathbf{M}_{j}$ has $\left|\mathbf{M}_{j}\right|$
detectors and a total of $\prod_{j=1}^{\ell}\left|\mathbf{M}_{j}\right|$
different paths can be observed over $\mathcal{S}_{\mathcal{M}}$.
From repeated experiments using the same setup, a table can then be
compiled that counts how often each allowable path occurred.

The probability of an individual path $A$ is defined as the conditional
probability
\begin{align}
P & :\mathcal{S}_{\mathcal{M}}\rightarrow\left[0,1\right]\in\mathbb{R},\nonumber \\
P\left(A\right) & :=\Pr\left(A_{\ell},A_{\ell-1},\ldots A_{2}\mid A_{1}\right).
\end{align}
Conditionalization on the first result $S\left(A\right)=A_{1}$ ensures
that the probability $P\left(A\right)$ is independent of the system's
history prior to the first measurement. We say that a path's source
establishes closure.

Note that we're only making statements about the probability of transitions
\emph{between} measurement outcomes, not about the probabilities of
outcomes by themselves. All measurement sequences are considered prepared
by the first measurement result, which yields conditional probabilities
for measurements thereafter.

\subsection{\protect\label{subsec:StrongRepeatabilityFactorization}Strong repeatability
and factorization}

Repeated measurements made from the identical detectors always produce
the same result\footnote{In this paper, we don't consider interactions (or any other dynamics)
to happen between measurements. Therefore, the wording ``repeated,''
``subsequent,'' or ``consecutive'' for measurements always implies
that nothing of consideration is happening in between.}. In a sequence $\mathcal{M}=\left[\mathbf{M}_{1},\mathbf{M}_{2},\ldots,\mathbf{M}_{\ell}\right]$
that includes two consecutive identical measurements, $\mathbf{M}_{j}=\mathbf{M}_{j+1}$,
the outcomes of $\mathbf{M}_{j}$ and $\mathbf{M}_{j+1}$ will always
be identical, no matter the preparations at $\mathbf{M}_{1}$. With
that, one of the $\mathbf{M}_{j},\mathbf{M}_{j+1}$ is redundant and
can be removed from the calculation entirely. We call this \emph{strong
repeatability}.

Paths must start and end with atomic measurements and have atomic
source and target outcomes. If a measurement sequence $\mathcal{M}^{\mathrm{A}}$
contains another atomic measurement, $^{\alpha}\mathbf{M}_{j}$ with
$j\neq1,\ell$, then we can decompose $\mathcal{M}^{\mathrm{A}}$
into two sequences, $\mathcal{M}^{\mathrm{B}}$ and $\mathcal{M}^{\mathrm{C}}$,
effectively duplicating $^{\alpha}\mathbf{M}_{j}$ and designating
it as $\mathbf{T}\left(B\right)$ and $\mathbf{S}\left(C\right)$,
respectively:
\begin{align}
\mathcal{M}^{\mathrm{A}} & =\left[\mathbf{M}_{1},\mathbf{M}_{2},\ldots,{}^{\alpha}\mathbf{M}_{j},\ldots,\mathbf{M}_{\ell}\right],\nonumber \\
\mathcal{M}^{\mathrm{B}} & :=\left[\mathbf{M}_{1},\mathbf{M}_{2},\ldots,{}^{\alpha}\mathbf{M}_{j}\right],\nonumber \\
\mathcal{M}^{\mathrm{C}} & :=\left[^{\alpha}\mathbf{M}_{j},\mathbf{M}_{j+1},\ldots,\mathbf{M}_{\ell}\right],\nonumber \\
\mathcal{M}^{\mathrm{A}} & =\mathcal{M}^{\mathrm{B}}\cdot\mathcal{M}^{\mathrm{C}}.
\end{align}
This applies to paths accordingly:
\begin{align}
A & \in\mathcal{S}_{\mathcal{M}^{\mathrm{A}}},\,\,\,\,B\in\mathcal{S}_{\mathcal{M}^{\mathrm{B}}},\,\,\,\,C\in\mathcal{S}_{\mathcal{M}^{\mathrm{C}}},\nonumber \\
A & =B\cdot C.
\end{align}
With this, sequences $\mathcal{M}^{\mathrm{A}}$ and $A$ can always
be unchained:
\begin{align}
\mathcal{M}^{\mathrm{B}} & =\mathcal{M}^{\mathrm{A}}/\mathcal{M}^{\mathrm{C}},\\
\mathcal{M}^{\mathrm{C}} & =\mathcal{M}^{\mathrm{B}}\setminus\mathcal{M}^{\mathrm{A}},\\
B & =A/C,\\
C & =B\setminus A.
\end{align}
This process of unchaining can be repeated on $\mathcal{M}^{\mathrm{A}}$
until all measurements $\mathcal{M}^{\mathrm{J}}$, $\mathrm{J}=1\ldots\mathrm{L}$,
and paths $A^{\mathrm{J}}$ are atomic only at their respective source
and target:
\begin{align}
\mathcal{M}^{\mathrm{A}} & =\prod_{\mathrm{J}=1}^{\mathrm{L}}\mathcal{M}^{\mathrm{J}}=\mathcal{M}^{1}\cdot\mathcal{M}^{2}\cdot\ldots\cdot\mathcal{M}^{\mathrm{L}},\\
A & =\prod_{\mathrm{J}=1}^{\mathrm{L}}A^{\mathrm{J}}=A^{1}\cdot A^{2}\cdot\ldots\cdot A^{\mathrm{L}}.
\end{align}
This process is called \emph{factorization}. The individual $\mathcal{M}^{\mathrm{J}}$
and $\mathcal{A}^{\mathrm{J}}$ are called undecomposable (or ununchainable,
if you wish).

\subsection{Closure as Markov property: Histories don't matter}

Since the probability of a path is independent of the history of the
system prior to its first measurement, factorization of paths corresponds
to the Markov property for conditional probabilities in conventional
probability theory,
\begin{align}
P\left(A\right) & =\Pr\left(A_{\ell_{\mathrm{L}}}^{\mathrm{L}},\ldots,A_{1}^{\mathrm{L}},A_{\ell_{\mathrm{L}-1}}^{\mathrm{L}-1},\ldots A_{1}^{\mathrm{L}-1},\ldots,A_{\ell_{2}}^{2},\ldots A_{1}^{2},A_{\ell_{1}}^{1},\ldots A_{2}^{1}\mid A_{1}^{1}\right).\nonumber \\
 & =\prod_{\mathrm{J}=1}^{\mathrm{L}}\Pr\left(A_{\ell_{\mathrm{J}}}^{\mathrm{J}},A_{\ell_{\mathrm{J}}-1}^{\mathrm{J}},\ldots A_{2}^{\mathrm{J}}\mid A_{1}^{\mathrm{J}}\right)=\prod_{\mathrm{J}=1}^{\mathrm{L}}P\left(A^{\mathrm{J}}\right).\label{eq:MarkovPropertyOfProbabilities}
\end{align}
We say that any atomic measurement establishes closure, such that
histories do not matter at that point.

\subsection{Weak repeatability}

While strong repeatability requires identical measurements, we now
also require weakly equivalent detectors to preserve the measurement
of underlying atomic outcome elements, whether observed as an atomic
result or assumed as part of a coarsened (non-atomic) result. This
is called \emph{weak repeatability}\footnote{This term was introduced in \cite{BCL1990} for a similar notion in
the context of ``generalized quantum mechanics'', which explored
generalization of an observable compatible with operator quantum mechanics
and quantum states.}.
\begin{figure}
\begin{centering}
\includegraphics[viewport=140bp 40bp 1060bp 800bp,clip,width=15cm]{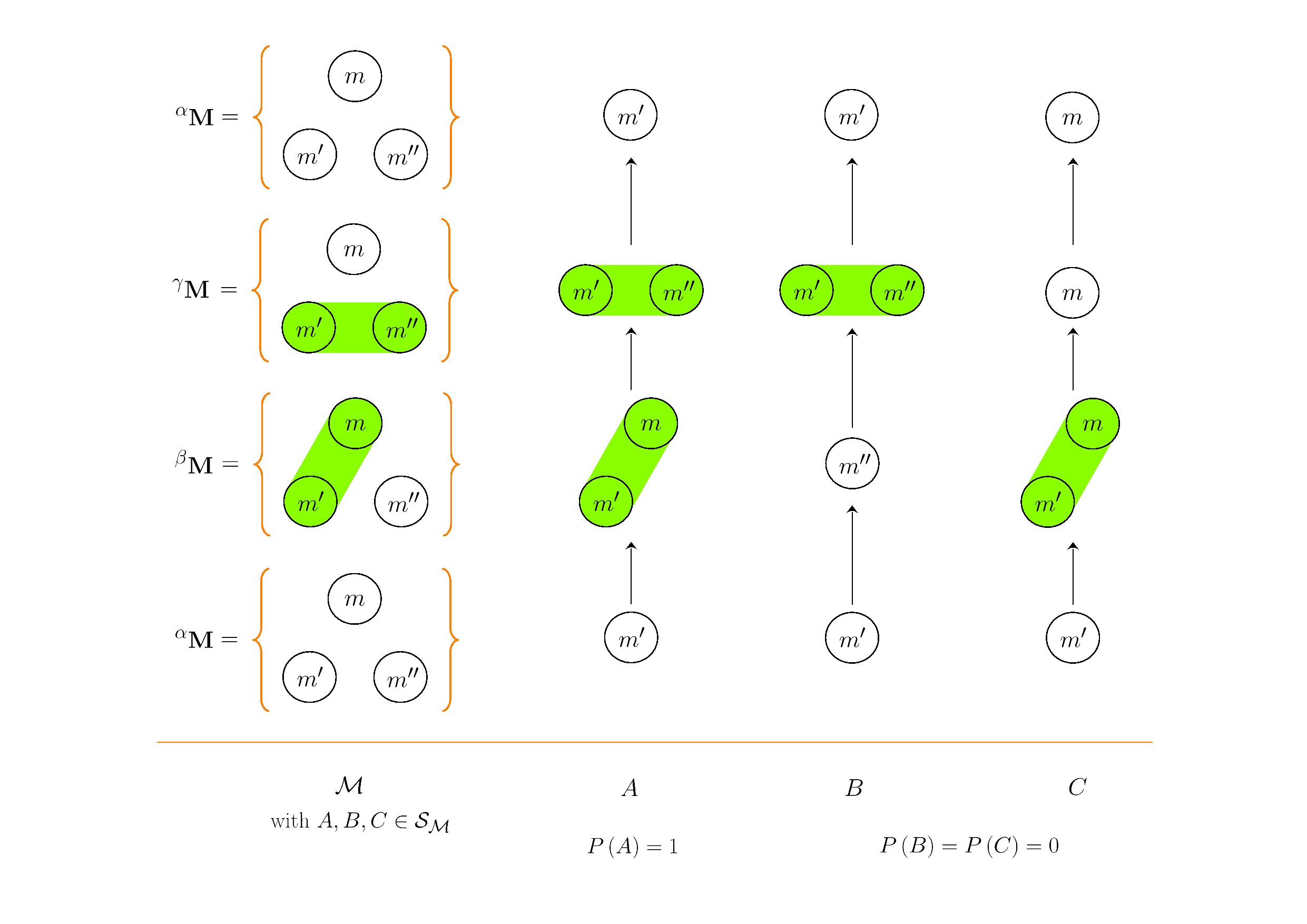}
\par\end{centering}
\caption{\protect\label{fig:weakRepeatabilityExample}In the example of an
experiment $\mathcal{M}=\left[^{\alpha}\mathbf{M},{}^{\beta}\mathbf{M},{}^{\gamma}\mathbf{M},{}^{\alpha}\mathbf{M}\right],$
made from three weakly equivalent detectors $^{\alpha}\mathbf{M}\sim{}^{\beta}\mathbf{M}\sim{}^{\gamma}\mathbf{M}$,
the probability of a sequence of measurement results is $1$ if all
detectors share the same underlying atomic outcome element (here:
$m^{\prime}$ is part of all detectors of $A$); otherwise, the probability
is $0$.}
\end{figure}

For example (figure \ref{fig:weakRepeatabilityExample}), the measurements
\begin{align}
^{\alpha}\mathbf{M} & =\left\{ \left\{ m\right\} ,\left\{ m^{\prime}\right\} ,\left\{ m^{\prime\prime}\right\} \right\} ,\nonumber \\
^{\beta}\mathbf{M} & =\left\{ \left\{ m,m^{\prime}\right\} ,\left\{ m^{\prime\prime}\right\} \right\} ,\\
^{\gamma}\mathbf{M} & =\left\{ \left\{ m\right\} ,\left\{ m^{\prime},m^{\prime\prime}\right\} \right\} ,\nonumber 
\end{align}
are weakly equivalent, $^{\alpha}\mathbf{M}\sim{}^{\beta}\mathbf{M}\sim{}^{\gamma}\mathbf{M}$,
since they share the same underlying atomic outcome elements $m,m^{\prime},m^{\prime\prime}$.
Weak repeatability then preserves the overlapping measurement outcome
$m^{\prime}$ in the path $A\in\mathcal{S_{M}}$:
\begin{align}
\mathcal{M} & =\left[^{\alpha}\mathbf{M},{}^{\beta}\mathbf{M},{}^{\gamma}\mathbf{M},{}^{\alpha}\mathbf{M}\right],\\
A & =\left[\left\{ m^{\prime}\right\} ,\left\{ m,m^{\prime}\right\} ,\left\{ m^{\prime},m^{\prime\prime}\right\} ,\left\{ m^{\prime}\right\} \right].
\end{align}

Note that $A$ is the only path with nonzero probability over $\mathcal{M}$
that starts with $\left\{ m^{\prime}\right\} $ since the underlying
atomic outcome element $m^{\prime}$ is preserved in the experiment.
All other paths starting with $\left\{ m^{\prime}\right\} $ have
probability zero and are hence impossible, e.g. $B,C\in\mathcal{S_{M}}$,
\begin{align}
B & =\left[\left\{ m^{\prime}\right\} ,\left\{ m^{\prime\prime}\right\} ,\left\{ m^{\prime},m^{\prime\prime}\right\} ,\left\{ m^{\prime}\right\} \right]\textrm{, or}\\
C & =\left[\left\{ m^{\prime}\right\} ,\left\{ m,m^{\prime}\right\} ,\left\{ m\right\} ,\left\{ m\right\} \right].
\end{align}
Path $B$ is impossible because $m^{\prime}$ is not contained in
the outcome $\left\{ m^{\prime\prime}\right\} \in{}^{\beta}\mathbf{M}$,
and path $C$ is impossible because $m^{\prime}$ is not contained
in $\left\{ m\right\} \in{}^{\gamma}\mathbf{M}$ (it does not matter
that $m$ is in $\left\{ m,m^{\prime}\right\} \in{}^{\beta}\mathbf{M}$).
This gives us probabilities for these paths:
\begin{align}
P\left(A\right) & =\Pr\left(\left\{ m^{\prime}\right\} ,\left\{ m^{\prime},m^{\prime\prime}\right\} ,\left\{ m,m^{\prime}\right\} \mid\left\{ m^{\prime}\right\} \right)=1,\\
P\left(B\right)=P\left(C\right) & =0.
\end{align}

Weak repeatability also allows inference given coarse-grained measurements:
In an arbitrary sequence that somewhere contains $^{\beta}\mathbf{M}$
directly followed by $^{\alpha}\mathbf{M}$, once outcome $\left\{ m,m^{\prime}\right\} \in{}^{\beta}\mathbf{M}$
is detected, the subsequent measurement $^{\alpha}\mathbf{M}$ will
yield one of $\left\{ m\right\} $ or $\left\{ m^{\prime}\right\} $
with certainty (and never $\left\{ m^{\prime\prime}\right\} $). Note
that this inference applies only to consecutive weakly equivalent
measurements.

\subsection{Certain measurements\protect\label{subsec:Certain-measurements}}

Measurements consisting of only a single detector will always measure
the outcome from that detector. Such a detector is fully coarse-grained
across all underlying atomic outcome elements, and the measurement
result is certain.

Certain measurements\footnote{The ``no disturbance'' postulate in \cite{Goyal2014} used the term
``trivial measurement'' for the same concept. In this paper, we
chose ``certain'' here to avoid confusion with our use of ``trivial''
for repeated identical measurements.} may be inserted anywhere into any existing path without changing
that path's probability. Given any $\mathcal{M}\in\left[\mathbb{M}\right]$,
$A\in\mathcal{S}_{\mathcal{M}}$, a fully coarse-grained measurement
$^{\cup}\mathbf{M}\in\mathbb{M}$ with single outcome, $^{\cup}\mathbf{M}=\left\{ ^{\cup}M\right\} $,
yields a certain measurement result, and we have for any $j\in\left\{ 2,\ldots,\ell_{\mathcal{M}}\right\} $:
\begin{align}
P\left(A\downarrow_{j}{}^{\cup}M\right) & =P\left(A\right).
\end{align}

\subsection{\protect\label{subsec:Impossible-paths}Impossible paths}

Paths that contain different results from two subsequent identical
measurements are impossible due to strong repeatability, i.e., they
have a probability of $0$. Likewise, paths with two subsequent weakly
equivalent measurements are impossible if they do not share at least
one underlying atomic outcome element in their respective results.

In both cases, the impossible paths contain at least one \emph{impossibility-generating
pair} (or short, IGP), a pair of consecutive weakly equivalent measurements,
where the results do not contain a common possible atomic outcome
element.

We now introduce impossible paths as a formal construct that is not
realized in nature and derive specific properties in the model algebra:
\begin{enumerate}
\item Two impossible paths coarsened (or refined) always result in an impossible
path,
\item coarsening of a possible path with an impossible path always yields
a possible path,
\item refinement of a possible path by an impossible path always yields
a possible path.
\end{enumerate}
Note that impossible paths are formally valid in the model algebra
and operations thereon; they merely have \emph{probability} zero.
This is distinct from operands in disallowed operations, which cannot
be executed in the partial model algebra in principle.

First, to prove that two impossible paths coarsened (or refined) must
always yield an impossible path (``1.''):
\begin{proof}
Coarsening may only modify exactly one element in a measurement sequence.
There are two cases: (1) the sequence step modified under coarsening
is not part of an IGP, or (2) the sequence step is one of an IGP.
In case (1), the resulting path must also be impossible, since no
IGP is modified. In the case of (2), only one of the two measurements in the IGP is modified.
If the result would make the IGP possible, this would mean that the
resulting path contains the same underlying atomic outcome element
in both measurement results of the original IGP. This would be a contradiction
because only one of the results is modified, and both initial paths
are impossible and therefore do not contain any shared underlying
atomic outcome elements at the IGP. A similar argument applies to
refinement.
\end{proof}
Now proving that coarsening a possible path with an impossible path
always yields a possible path (``2.''):
\begin{proof}
Coarsening two paths always results in a path that has a single sequence
step modified. There, the coarsened result is built over the initial
paths' underlying atomic outcome elements. Because this always includes
all outcome elements from the initial paths, it will contain all elements
from the possible path.
\end{proof}
Finally, the proof that refining a possible path by an impossible
path always yields another possible path (``3.''):
\begin{proof}
Refinement of a possible path by an impossible path must always happen
at the single IGP of the impossible path since, otherwise, both paths
would contain that IGP and be impossible. At the IGP, the possible
and impossible paths have one measurement result in common and differ
by the other. At those two steps, the possible path must share at
least one underlying atomic outcome element across both measurement
results. If the impossible path were to remove these shared outcome
elements from the possible path, the impossible path would have also
had to contain these outcome elements in the IGP. This would make
the impossible path possible and is a contradiction.
\end{proof}

\subsection{\protect\label{subsec:Equivalent-paths-Redundancy}Equivalent paths
and redundancy}

Per strong repeatability, one of two consecutive equal measurements
may be removed from the calculation. Duplicating measurements could,
therefore, be viewed as some identity operation under chaining in
the model algebra. Similarly, elements from the space of impossible
paths act on other paths as identity elements under coarsening.

We now formally define a class of equivalent paths from these two
scenarios, such that any two equivalent paths can be exchanged at
will, and treated as equal in the calculations to come.

Two paths are equivalent if they can be changed into one another \ldots{}
\begin{enumerate}
\item \ldots{} through coarsening or refinement with impossible paths; or
\item \ldots{} through chaining, unchaining, insertion, or removal of identical
measurement results from consecutive equal measurements.
\end{enumerate}
Given all possible paths that are equivalent under this class, there
is a unique minimal possible path, now called \emph{nonredundant},
such that every other path equivalent to it contains some redundancy.
\begin{proof}
Per (1), coarsening and refinement are generally only allowed if the
operands differ by one measurement result at a single sequence step.
At all other sequence steps, the paths must be identical. Coarsening
a possible path with an impossible path, therefore, enlarges the possible
path by outcome elements of an IGP. Repeated refinement, in turn,
reduces the number of elements from a path. Once all elements from
any IGP are removed from a possible path, such a path is uniquely
minimal under refinement. Per (2), identical measurement results can
be trivially chained, unchained, inserted, or removed. A path is uniquely
minimal under removal once it contains no more consecutive duplicates.
Taken together, unchaining from (1) and removal from (2) can be repeated
on any given possible path until it is uniquely minimal in both cases.
The resulting path is, therefore, possible, unique under the class,
and contains no redundancy.
\end{proof}
We extend the definition of symmetric (palindromic) paths accordingly
to paths that are symmetric in their nonredundant form.

For notation, two equivalent paths $A,B$ may be written $A\sim B$
when emphasizing the possibility of different redundancy in $A$ and
$B$; but generally, $A=B$ suffices since equivalent paths are effectively
equal.

\subsection{Extending coarsening for equivalent paths}

When we first introduced coarsening in section \ref{subsec:Chaining-and-coarsening},
we defined $\mathcal{M}^{\vee\left\{ j\right\} }$ as the sequence
that differs from a given sequence $\mathcal{M}$ by replacing exactly
one of its measurements, $\mathbf{M}_{j}$, with a coarsened version
$^{\vee}\mathbf{M}_{j}$. The coarsening operation $A\vee B=C$ is
then allowable only if $A,B\in\mathcal{S}_{\mathcal{M}}$ and $C\in\mathcal{S}_{\mathcal{M}^{\lor\left\{ j\right\} }}$,
and the underlying atomic outcome elements of $A$ and $B$ at $j$
do not overlap.

Most general paths do not meet these conditions and cannot be coarsened.
However, even if these conditions are met for specific paths $A,B,C$,
they are generally not satisfied for paths equivalent to these under
the redundancy class. Therefore, we extend the partial operation coarsening
$A\vee B=C$ to be allowable if the operation's prerequisites can
be met for at least one equivalent member representing each of the
operands. This is consistent with treating equivalent paths as equal
while evaluating expressions in the model algebra on suitable members
representing that equivalence class.

\subsection{\protect\label{subsec:Connection-to-classical}Connection to classical
probability}

Insertion of a certain measurement does not change a path's probability,
whereas insertion of a measurement that generates a new IGP forces
the path's probability to $0$. These operations are conceptually
similar to conventional probability, where insertion of a certain
or impossible event into an existing set of independent events has
the same effect on the joint probability for these events\footnote{Note that insertion of a certain measurement into a path does not
have to generate redundancy and is therefore not necessarily trivial,
as it would be in conventional probability.}.

Furthermore, in an experiment with two measurements, $\mathcal{M}^{\mathbf{D}}=\left[\mathbf{M}_{1},\mathbf{M}_{2}\right]\in\left[\mathbb{M}\right]$,
the sum of probabilities over all possible paths $D_{r}$ that start
with $M_{1}\in\mathbf{M}_{1}$ and end on one of the $M_{2}^{r}\in\mathbf{M}_{2}$
must add up to $1$, since detectors are distinct and non-filtering:
\begin{align}
D_{r} & :=\left[M_{1},M_{2}^{r}\right]\in\mathcal{S}_{\mathcal{M^{\mathbf{D}}}},\\
\sum_{\forall r}P\left(D_{r}\right) & =\sum_{\forall r}\Pr\left(M_{2}^{r}\mid M_{1}\right)=1.
\end{align}

\section{\protect\label{sec:Amplitude-algebra}Amplitude algebra}

Equipped with an axiomatic set of partial operations from an elementary
model and specific questions we want to ask from that model, we can
now choose the concept of \emph{amplitudes} to represent answers to
our questions. We will support this choice with a simplicity argument,
show that it is valid without loss of generality for the task at hand,
and not in conflict with the model in general. This executes our stated,
programmatic goal from the introduction: to avoid premature optimization
of the axiomatic model algebra, for which we may not have enough structure
to attempt classification; to bring forward a choice that allows answering
specific questions; and to provide the basis for complete classification
of consequences from that choice, including an explanation towards
the utility of the complex numbers in canonical Quantum Mechanics.

\subsection{\protect\label{subsec:RepresentationSpace}Representation space}

An amplitude $\mathbf{a}$ is an element in an axiomatic space $\mathbb{A}$
that is now constructed to quantify probability behavior under operations
of the model algebra. Given an experiment $\mathcal{M}\in\left[\mathbb{M}\right]$
and path $A\in\mathcal{S}_{\mathcal{M}}$, a function $\phi$ represents
that path $A$ as an amplitude $\mathbf{a}\in\mathbb{A}$:
\begin{align}
\phi & :\mathcal{S}_{\left[\mathbb{M}\right]}\rightarrow\mathbb{A},\nonumber \\
\mathcal{M} & \in\left[\mathbb{M}\right],\,\,\,\,A\in\mathcal{S}_{\mathcal{M}},\,\,\,\,\mathbf{a}\in\mathbb{A},\\
\mathbf{a} & =\phi\left(A\right).\nonumber 
\end{align}
The representation function $\phi$ is a well-defined map of outcomes
of an experiment $\mathcal{S}_{\left[\mathbb{M}\right]}$ into the
space of amplitudes $\mathbb{A}$. Equivalent paths $A,B\in\mathcal{S}_{\mathcal{M}}$
have identical representation,
\begin{equation}
A\sim B\quad\rightarrow\quad\phi\left(A\right)=\phi\left(B\right).
\end{equation}
The representation map $\phi$ does not have to be invertible, and
not every value in $\mathbb{A}$ has to represent an allowable path
in $\mathcal{S}_{\left[\mathbb{M}\right]}$.

The probability $P$ of a path is identical to the $p$-function of
the amplitude representing that path:
\begin{align}
p & :\mathbb{A}\rightarrow\mathbb{R},\nonumber \\
p\left(\phi\left(A\right)\right) & =P\left(A\right).
\end{align}
For amplitudes that do not represent allowable paths, the image of
the $p$-function is not bounded to the probability interval $\left[0,1\right]\in\mathbb{R}$
but formally extended to all of $\mathbb{R}$.

The concept of disallowed subspaces in amplitude representation of
nature is nothing new: In canonical Quantum Mechanics, distinct states
of some system may be represented as amplitudes in the space of complex
numbers, $\mathbb{C}$. There, the Born rule allows to compute the
probability of finding the system in one of these states, by forming
the squared norm of their respective amplitudes. Because probabilities
greater than $1$ are nonsensical, any number in $\mathbb{C}$ with
a norm greater than $1$ can never represent an allowable state in
these systems. However, all complex numbers are formally useful in
the calculation, if abstract.

By tracing algebraic operations on amplitudes in $\mathbb{A}$ from
a model algebra, regardless of whether such operations would make
sense in the original model, we make a subtle but important choice:
\begin{itemize}
\item Any valid operation in the model algebra \emph{must} have a well-defined
algebraic representation on amplitudes, and
\item any permissible operation on amplitudes \emph{must not} conflict with
allowable operations in the model.
\end{itemize}
These two constraints allow the freedom of valid algebraic properties
on amplitudes that cannot be traced from the model in principle. All
valid properties on amplitudes -- traceable or not -- must have
utility when answering questions the observer brings to the experiment,
namely, asking for probabilities.

\subsection{Core algebraic properties}

The model algebra of paths $\mathcal{S}_{\left[\mathbb{M}\right]}$
over experiments $\left[\mathbb{M}\right]$ has coarsening $\vee:\mathcal{S}_{\mathbb{\left[\mathbb{M}\right]}}\times\mathcal{S}_{\mathbb{\left[\mathbb{M}\right]}}\rightharpoonup\mathcal{S}_{\left[\mathbb{M}\right]}$
and chaining $\cdot:\mathcal{S}_{\left[\mathbb{M}\right]}\times\mathcal{S}_{\left[\mathbb{M}\right]}\rightharpoonup\mathcal{S}_{\left[\mathbb{M}\right]}$
as its elementary partial operations, abbreviated $\left\langle \mathcal{S}_{\left[\mathbb{M}\right]},\vee,\cdot\right\rangle $.
This is now paired with algebraic operations $\oplus$ and $\odot$
on the amplitude space, $\left\langle \mathbb{A},\oplus,\odot\right\rangle $
using some mapping $\phi$:
\begin{equation}
\left\langle \mathcal{S}_{\left[\mathbb{M}\right]},\vee,\cdot\right\rangle \overset{\phi}{\longmapsto}\left\langle \mathbb{A},\oplus,\odot\right\rangle .
\end{equation}
In contrast to the model algebra, binary operations in the amplitude
algebra are defined on all elements of the space. The symbols $\oplus$
and $\odot$ are chosen suggestively, termed amplitude addition and
multiplication, respectively. At this point, however, these are just
placeholder symbols and terms.

For some experiment $\mathcal{M}\in\left[\mathbb{M}\right]$ and paths
$A,B\in\mathcal{S}_{\mathcal{M}}$ that are allowed to be coarsened,
the amplitude mapping $\phi:\mathcal{S}_{\left[\mathbb{M}\right]}\rightarrow\mathbb{A}$
has the property:
\begin{equation}
\phi\left(A\vee B\right)=\phi\left(A\right)\oplus\phi\left(B\right).
\end{equation}
Coarsening is commutative and partially associative, from which we
trace for any $\mathbf{a},\mathbf{b},\mathbf{c}\in\mathbb{A}$:
\begin{align}
\oplus & :\mathbb{A}\times\mathbb{A}\rightarrow\mathbb{A},\nonumber \\
\left(\mathbf{a}\oplus\mathbf{b}\right)\oplus\mathbf{c} & =\mathbf{a}\oplus\left(\mathbf{b}\oplus\mathbf{c}\right),\\
\mathbf{a}\oplus\mathbf{b} & =\mathbf{b}\oplus\mathbf{a}.
\end{align}
For chaining, we have accordingly:
\begin{equation}
\phi\left(A\cdot B\right)=\phi\left(A\right)\odot\phi\left(B\right).
\end{equation}
Closure as Markov property for probabilities (equation \ref{eq:MarkovPropertyOfProbabilities})
extends to amplitudes:
\begin{align}
P\left(A\cdot B\right) & =P\left(A\right)P\left(B\right)\nonumber \\
=p\left(\phi\left(A\cdot B\right)\right) & =p\left(\phi\left(A\right)\odot\phi\left(B\right)\right)\\
=p\left(\mathbf{a}\odot\mathbf{b}\right) & =p\left(\mathbf{a}\right)p\left(\mathbf{b}\right).\nonumber 
\end{align}

Chaining is not required to be commutative but associative. This becomes:
\begin{align}
\odot & :\mathbb{A}\times\mathbb{A}\rightarrow\mathbb{A},\nonumber \\
\left(\mathbf{a}\odot\mathbf{b}\right)\odot\mathbf{c} & =\mathbf{a}\odot\left(\mathbf{b}\odot\mathbf{c}\right).
\end{align}
Chaining distributes over coarsening, which holds in amplitude representation
as well:
\begin{align}
\left(\mathbf{a}\oplus\mathbf{b}\right)\odot\mathbf{c} & =\mathbf{a}\odot\mathbf{c}\oplus\mathbf{b}\odot\mathbf{c},\\
\mathbf{a}\odot\left(\mathbf{b}\oplus\mathbf{c}\right) & =\mathbf{a}\odot\mathbf{b}\oplus\mathbf{a}\odot\mathbf{c}.
\end{align}
These are the core algebraic relations (S1) through (S5) from \cite{GKS2010}.

\subsection{Involution}

Path reversal $\overline{\phantom{{A}}}:\mathbb{\left[\mathbb{M}\right]}\rightarrow\mathbb{\left[\mathbb{M}\right]}$
is an involution that is an anti-isomorphism, i.e., it distributes
over coarsening, $\vee$, but is an anti-involution over chaining,
$\cdot$. This traces to representation in the amplitude algebra:
\begin{align}
\overline{\phantom{{A}}} & :\mathbb{A}\rightarrow\mathbb{A},\nonumber \\
\overline{\overline{\mathbf{a}}} & =\mathbf{a},\\
\overline{\mathbf{a}\oplus\mathbf{b}}=\overline{\mathbf{b}\oplus\mathbf{a}} & =\overline{\mathbf{a}}\oplus\overline{\mathbf{b}}=\overline{\mathbf{b}}\oplus\overline{\mathbf{a}},\\
\overline{\mathbf{a}\odot\mathbf{b}} & =\overline{\mathbf{b}}\odot\overline{\mathbf{a}}.
\end{align}
Self-adjoint paths $\,^{\circledcirc}\negthinspace A$ are represented
by amplitudes $\mathbf{d}=\phi\left(\,^{\circledcirc}\negthinspace A\right)$
that are invariant under involution, $\overline{\mathbf{d}}=\mathbf{d}$.

\subsection{\protect\label{subsec:ImpossiblePathsAsAdditiveIdentity}Impossible
paths as additive identity}

Section \ref{subsec:Impossible-paths} introduced impossible paths
as a formal construct for sequences of measurement results with probability
$0$. They contain at least one pair of results from weakly equivalent
measurements that do not share an underlying atomic outcome element.
We observe that impossible paths as a whole have properties similar
to an additive identity under coarsening: Coarsening two impossible
paths always results in another impossible path, coarsening a possible
path with an impossible path always yields a possible path, and coarsening
two possible paths always yields a possible path.

We, therefore, now choose to represent all impossible paths by the
same amplitude, $\mathbf{0}\in\mathbb{A}$, and to be the unique additive
identity on amplitudes. Given an arbitrary possible path $A$ with
$\mathbf{a}=\phi\left(A\right)$, there is:
\begin{align}
\mathbf{0}\oplus\mathbf{0} & =\mathbf{0},\\
\mathbf{0}\oplus\mathbf{a}=\mathbf{a}\oplus\mathbf{0} & =\mathbf{a}.
\end{align}
The probability for any impossible path is $0$, which applies to
their amplitude representation as well:
\begin{align}
p\left(\mathbf{0}\right) & =0.
\end{align}
Paths that are possible, in contrast, have nonzero probability. For
allowable paths path $B$ with $\mathbf{b}=\phi\left(B\right)$, we
can therefore infer
\begin{align}
p\left(\phi\left(B\right)\right) & =0\quad\rightarrow\quad\mathbf{b}=\mathbf{0}.
\end{align}
Note that we cannot make assertions about elements in $\mathbb{A}$
that do not represent allowable paths. If there are elements $\mathbf{c}\in\mathbb{A}$
where $p\left(\mathbf{c}\right)=0$ but $\mathbf{c}\neq\mathbf{0}$,
then such $\mathbf{c}$ must never represent allowable paths in the
model algebra.

Coarsening with an impossible path is a redundancy operation between
equivalent paths per section \ref{subsec:Equivalent-paths-Redundancy}.
This is consistent with the algebraic constraint imposed on amplitudes
here: Equivalent paths are exchangeable in the calculation at will
and can therefore be represented by the same amplitude, $\mathbf{0}$.

\subsection{Trivial paths as multiplicative identity}

A trivial path is a sequence of results from an experiment that consists
only of equal measurements. Per strong repeatability, all measurement
results must be identical. Like coarsening with an impossible path,
chaining with a trivial path is a redundancy operation.

We formally choose the model operation of chaining with a trivial
path to be represented by an amplitude, $\mathbf{1}\in\mathbb{A}$,
such that for any path $A$ with $\mathbf{a}=\phi\left(A\right)$,
there is:
\begin{align}
\mathbf{a}\odot\mathbf{1}=\mathbf{1}\odot\mathbf{a} & =\mathbf{a}.
\end{align}
Because $\mathbf{1}$ is a left- and right-identity, and amplitude
multiplication $\odot$ is associative, $\mathbf{1}$ is a multiplicative
identity for all $\mathbf{a}\in\mathbb{A}$, and
\begin{equation}
p\left(\mathbf{1}\right)=1.
\end{equation}

In the model algebra, a trivial path $\,^{\mathbf{1}}\negthinspace A$
can be chained to any of the $A^{\mathrm{J}}$ of a path factorization
$A=\prod_{\mathrm{J}=1}^{\mathrm{L}}A^{\mathrm{J}}=A^{1}\cdot A^{2}\cdot\ldots\cdot A^{\mathrm{L}}$,
as long as all measurements in $\,^{\mathbf{1}}\negthinspace A$ are
the same as the target of $A^{\mathrm{J}}$ (which is also the source
of $A^{\mathrm{J}+1}$, of course):
\begin{align}
A^{\prime} & =A^{1}\cdot\ldots A^{\mathrm{J}}\cdot\,^{\mathbf{1}}\negthinspace A\cdot A^{\mathrm{J}+1}\ldots\cdot A^{\mathrm{L}},\\
\phi\left(A^{\prime}\right) & =\phi\left(A^{1}\right)\odot\ldots\odot\phi\left(A^{\mathrm{J}}\right)\odot\mathbf{1}\odot\phi\left(A^{\mathrm{J}+1}\right)\odot\ldots\odot\phi\left(A^{\mathrm{L}}\right)=\phi\left(A\right).
\end{align}
This correspond to insertion of $\,^{\mathbf{1}}\negthinspace A$,
\begin{equation}
A^{\prime}=A\downarrow_{\mathrm{J}}\,^{\mathbf{1}}\negthinspace A.
\end{equation}

Because chaining with a trivial path is a redundancy operation, the
paths $A^{\prime}$ and $A$ are equal, and the length of their nonredundant
form is not changed. This is consistent with representation by a multiplicative
unit $\mathbf{1}=\phi\left(\,^{\mathbf{1}}\negthinspace A\right)$.
It also means that the insertion index $\mathrm{J}$ in the expression
$A\downarrow_{\mathrm{J}}\,^{\mathbf{1}}\negthinspace A$ is superfluous:
Wherever it is allowable to insert $\,^{\mathbf{1}}\negthinspace A$,
it results in an equivalent path. We therefore write:
\begin{equation}
A\sim A\downarrow\,^{\mathbf{1}}\negthinspace A.
\end{equation}

In contrast, note that insertions of a single, certain \emph{measurement}
$A\downarrow_{j}{}^{\cup}M$, as in section \ref{subsec:Certain-measurements},
do not correspond to chaining in the model algebra and, therefore,
have no algebraic representation. Even though this operation does
not change the probability of the path, $P\left(A\downarrow_{j}{}^{\cup}M\right)=P\left(A\right)$,
its amplitude $\phi\left(A\right)$ may be different from $\phi\left(A\downarrow_{j}{}^{\cup}M\right)$.

\subsection{Invertibility and algebraic closure}

What led to a coarsened path can be refined, and what led to a chained
path can be unchained. We argued for partially well-defined inverse
operations in the model algebra in section \ref{subsec:Inverse-operations-unchaining},
where we have complete information on all detectors, their possible
measurement outcomes, and actual measurements. In this section, we
now show that additive invertibility must always hold in amplitude
representation and that multiplicative invertibility must hold for
paths with nonzero probability. We postulate algebraic closure for
the formal existence of inverse elements in $\mathbb{A}$ that represent
these inverse operations.

\subsubsection{Additive inverse}

For every nonzero element $\mathbf{a}\in\mathbb{A}$, $\mathbf{a}\neq\mathbf{0}$
that represents possible path $A\in\mathcal{S}_{\mathcal{M}}$, there
must be an operation $\ominus$ that is inverse to $\oplus$ such
that
\begin{equation}
\left(\mathbf{0}\oplus\mathbf{a}\right)\ominus\mathbf{a}=\mathbf{0}.
\end{equation}
This follows directly from coarsening and subsequent refinement in
the model algebra (where $\left(A\vee B\right)\wedge B=A$ per equation
\ref{eq:refinementDef}) and from $p\left(\left(\mathbf{0}\oplus\mathbf{a}\right)\ominus\mathbf{a}\right)=0$.

Next to the existence of an inverse operation to addition, algebraic
closure requires the formal existence of inverse elements $-\mathbf{a}\in\mathbb{A}$
representing this operation, such that $\mathbf{a}\oplus\left(-\mathbf{a}\right)=\mathbf{0}$
and therefore $p\left(\mathbf{a}\oplus\left(-\mathbf{a}\right)\right)=0$.
Note that neither $-\mathbf{a}$ nor the expression $\mathbf{a}\oplus\left(-\mathbf{a}\right)$
represent paths or path operations from the model algebra anymore;
instead, these expressions are merely formally valid in amplitude
representation in a way that abstracts some earlier expressions that
did -- at first -- trace from path operations.

For testing uniqueness, we assume there exists a different amplitude
$\mathbf{a}^{\prime}\in\mathbb{A}$ such that $\mathbf{a}\neq\mathbf{a}^{\prime}$
but $\left(\mathbf{0}\oplus\mathbf{a}\right)\ominus\mathbf{a}^{\prime}=\mathbf{0}$
as well. However, since $\mathbf{0}\oplus\mathbf{a}$ represents a
possible path, refinement by another path can only yield an impossible
path if both are equivalent up to redundancy and therefore have the
same amplitude. Since that is not the case, $\left(\mathbf{0}\oplus\mathbf{a}\right)\ominus\mathbf{a}^{\prime}$
can never be zero for $\mathbf{a}\neq\mathbf{0}$ and $\mathbf{a}\neq\mathbf{a}^{\prime}$.
Because amplitude representation must not create expressions that
conflict with the model, such assumed $\mathbf{a}^{\prime}$ could
not exist, and $\ominus\mathbf{a}$ is unique for any given nonzero
$\mathbf{a}$.

The edge case $\mathbf{a}=\mathbf{a}^{\prime}=\mathbf{0}$ has $\left(\mathbf{0}\oplus\mathbf{0}\right)\ominus\mathbf{0}=\mathbf{0}$,
which is always true. Coarsening and refinement between impossible
paths always yield another impossible path. The unique additive inverse
for $\mathbf{a}=\mathbf{0}$, therefore, is $\mathbf{0}$ itself.

Existence, uniqueness, and algebraic closure of a two-sided
associative cancellation operation with identity element make amplitude
addition invertible in $\mathbb{A}$ for any $\mathbf{a}\in\mathbb{A}$.

\subsubsection{\protect\label{subsec:MultiplicativeInverse}Multiplicative inverse}

Similar to addition, multiplicative inverses exist as long as the
probability of the amplitude is nonzero. Given a possible path $A\in\mathcal{S}_{\mathcal{M}}$,
$\mathbf{a}\in\mathbb{A}$, $\phi\left(A\right)=\mathbf{a}\neq\mathbf{0}$,
there exist left- and right-inverse operations $\obslash$ and $\oslash$
to $\odot$ such that
\begin{align}
\left(\mathbf{1}\odot\mathbf{a}\right)\oslash\mathbf{a} & =\mathbf{1},\\
\mathbf{a}\obslash\left(\mathbf{a}\odot\mathbf{1}\right) & =\mathbf{1}.
\end{align}
This follows from chaining and subsequent left- and right-unchaining
in the model algebra (where $\left(A\cdot B\right)/B=A$ per equation
\ref{eq:rightUnchainingDef}, and $A\setminus\left(A\cdot B\right)=B$
per equation \ref{eq:leftUnchainingDef}).

Algebraic closure then requires the formal existence of inverse elements
$^{-1}\mathbf{a},\mathbf{a}^{-1}\in\mathbb{A}$ such that
\begin{align}
\left(\mathbf{1}\odot\mathbf{a}\right)\odot\mathbf{a}^{-1} & =\mathbf{1},\\
^{-1}\mathbf{a}\odot\left(\mathbf{a}\odot\mathbf{1}\right) & =\mathbf{1}.
\end{align}
Since $\mathbf{1}\odot\mathbf{a}=\mathbf{a}\odot\mathbf{1}$ and $\odot$
is associative, these left- and right-inverse operations must be the
same and can formally be represented by the same element under algebraic
closure,
\begin{equation}
^{-1}\mathbf{a}=\mathbf{a}^{-1}.
\end{equation}
Closure in the model algebra yields the Markov property for probability
(equation \ref{eq:MarkovPropertyOfProbabilities}), which for the
$p$-function formally extends to the closure space of $\mathbb{A}$
under amplitude multiplication:
\begin{align}
p\left(\mathbf{a}\odot\mathbf{a}^{-1}\right)=p\left(\mathbf{a}\right)p\left(\mathbf{a}^{-1}\right) & =1.
\end{align}
Just as the other expressions involving inverse elements, the $p$-function
of a multiplicative inverse is merely a formal extension of $p\left(\mathbf{a}\right)$
to elements of the closure space. The product composition property
$p\left(\mathbf{a}\odot\mathbf{b}\right)=p\left(\mathbf{a}\right)p\left(\mathbf{b}\right)$
is formally carried forward for inverses, $\mathbf{b}\equiv\mathbf{a}^{-1}$.
With this, for $p\left(\mathbf{a}\right)p\left(\mathbf{a}^{-1}\right)=1$
to hold and for inverses $\mathbf{a}^{-1}$ to exist, $p\left(\mathbf{a}\right)$
and $p\left(\mathbf{a}^{-1}\right)$ must be nonzero.

In order to argue for uniqueness, we assume some nonzero $\mathbf{a}^{\prime}\in\mathbb{A}$,
$\mathbf{a}^{\prime}\neq\mathbf{0}$ such that $\mathbf{a}\neq\mathbf{a}^{\prime}$
but $\left(\mathbf{1}\odot\mathbf{a}\right)\oslash\mathbf{a}^{\prime}=\mathbf{1}$.
Just as in the additive case, $\mathbf{a}^{\prime}\neq\mathbf{a}$
must correspond to a path $A^{\prime}$ that differs from $A$ in
its nonredundant form. Given a path $^{\mathbf{1}}B$ with unit amplitude
$\phi\left(^{\mathbf{1}}B\right)=\mathbf{1}$, unchaining $\left(^{\mathbf{1}}B\cdot A\right)/A^{\prime}$
requires the factorization of $A$ to end on the factorization of
$\overline{A^{\prime}}$, up to equivalence. For $\mathbf{a}^{\prime}\neq\mathbf{a}$
to hold, $A$ must, therefore, be longer than $A^{\prime}$, with
a remainder $\delta A=A/A^{\prime}$ and $\phi\left(\delta A\right)\neq\mathbf{1}$.
However, since $\phi\left(\delta A\right)=\phi\left(A/A^{\prime}\right)=\mathbf{a}\oslash\mathbf{a}^{\prime}\neq\mathbf{1}$,
this contradicts the requirement $\left(\mathbf{1}\odot\mathbf{a}\right)\oslash\mathbf{a}^{\prime}=\mathbf{1}$,
and this case is disallowed. The same argument applies to left-unchaining.
Just as for addition, expressions in amplitude representation must
never create valid expressions that conflict with the model. This
makes multiplicative inverses of amplitudes representing allowable
paths unique as well.

Therefore, we established the existence, uniqueness, and algebraic
closure of multiplicative inverses of amplitudes $\mathbf{a}\in\mathbb{A}$
from a two-sided associative cancellation operation with identity element,
as long as $p\left(\mathbf{a}\right)\neq0$.

\subsection{Scalars}

We now postulate the existence of a real-valued subspace of the amplitude
algebra that is spanned by the multiplicative identity amplitude $\mathbf{1}$,
\begin{equation}
\mathbb{R}\cong\mathbb{R}\mathbf{1}\subset\mathbb{A}.
\end{equation}
Elements $\lambda\in\mathbb{R}$ are called \emph{scalars} and exist
formally separate from amplitudes. We support this choice from conceptual
simplicity: The probabilities we are ultimately looking for are quantified
using real numbers between $0$ and $1$. In this paper, the mathematical
mechanism to obtain probabilities uses the more faceted construct
of amplitudes. By embedding the reals into these amplitudes by postulate,
we aim to keep amplitudes somewhat similar to probabilities, while
exposing additional facets as algebraic variations thereof.

Subsequently, we postulate for amplitudes $\lambda\mathbf{1}\in\mathbb{R}\mathbf{1}$
to represent symmetric (palindromic) paths in the model. This choice
is supported by internal consistency, where we will show that resulting
properties in the amplitude algebra either trace from the model or
do not conflict with it.

Whether these postulates are good choices cannot be answered mathematically.
Instead, their utility has to be shown in the experiment. You could
pointedly say that here we are doing physics.

Two symmetric paths $\,^{\circledcirc}\negthinspace B,\,^{\circledcirc}C\in\mathcal{S}_{^{\circledcirc}\negthinspace\mathcal{M}}$
with amplitudes $\mathbf{b}=\phi\left(\,^{\circledcirc}\negthinspace B\right),\mathbf{c}=\phi\left(\,^{\circledcirc}C\right)$,
\begin{align}
\overline{\,^{\circledcirc}\negthinspace B} & =\,^{\circledcirc}\negthinspace B, & \overline{\,^{\circledcirc}C} & =\,^{\circledcirc}C,\\
\overline{\mathbf{b}} & =\mathbf{b}, & \overline{\mathbf{c}} & =\mathbf{c},
\end{align}
correspond to real scalars $\lambda,\mu\in\mathbb{R}$:
\begin{align}
\mathbf{b} & :=\lambda\mathbf{1}\equiv\lambda, & \mathbf{c} & :=\mu\mathbf{1}\equiv\mu.
\end{align}
Since scalars formally exist as separate entities from amplitudes,
they may be notated separately and pulled up front by convention.

Chaining their paths then follows the same rules as for the remaining
multiplicative identity. For example, given some general path $A$
with $\mathbf{a}=\phi\left(A\right)$, chaining with paths representing
scalars is an insertion wherever allowable:
\begin{align}
\phi\left(A\downarrow\,^{\circledcirc}\negthinspace B\right) & =\lambda\,\phi\left(A\right),\\
\phi\left(A\downarrow\,^{\circledcirc}C\right) & =\mu\,\phi\left(A\right),\nonumber \\
\mathbf{a}\odot\mathbf{b}=\mathbf{a}\odot\lambda\mathbf{1} & =\lambda\mathbf{a}=\lambda\mathbf{1}\odot\mathbf{a}=\mathbf{b}\odot\mathbf{a},\\
\mathbf{a}\odot\mathbf{b}\odot\mathbf{c} & =\lambda\mu\mathbf{a}=\mathbf{b}\odot\mathbf{a}\odot\mathbf{c}=\ldots\textrm{(all permutations)}.
\end{align}
The zero scalar, $\nu=0$, is the annihilator under multiplication
that forces the probability of any sequence to $0$ and must therefore
effect a zero amplitude when inserted:
\begin{align}
0\mathbf{a} & =\mathbf{0},\\
\nu\mathbf{a} & =\mathbf{0}\textrm{ for any nonzero }\mathbf{a}\quad\rightarrow\quad\nu=0.
\end{align}

With commutativity given by postulate of the real numbers, we have:
\begin{equation}
\overline{\lambda\mathbf{1}\odot\mu\mathbf{1}}=\overline{\mu\mathbf{1}}\odot\overline{\lambda\mathbf{1}}=\mu\mathbf{1}\odot\lambda\mathbf{1}=\lambda\mathbf{1}\odot\mu\mathbf{1}.
\end{equation}
That is, the product of two amplitudes representing a symmetric path,
$\sigma:=\lambda\mu\equiv\lambda\mathbf{1}\odot\mu\mathbf{1}$, is
also self-adjoint, making the algebraic subspace closed under chaining.
This identifies self-adjoint amplitudes with the reals,
\begin{equation}
\left\{ \sigma\mid\sigma\in\mathbb{R}\right\} \cong\left\{ \mathbf{c}\mid\mathbf{c}\in\mathbb{A},\,\overline{\mathbf{c}}=\mathbf{c}\right\} ,\label{eq:selfAdjointAmplIsAlwaysScalar}
\end{equation}
independently of whether the represented path in the model is symmetric
or not\footnote{The case where an amplitude equal to its involution would not be represented
by a real number conflicts with the requirement in section \ref{subsec:RepresentationSpace}
that algebraic representations be well-defined.}.

Because involution is linear,
\begin{equation}
\overline{\lambda\mathbf{1}\odot\mathbf{a}}=\lambda\mathbf{1}\odot\overline{\mathbf{a}}\equiv\lambda\overline{\mathbf{a}}=\overline{\lambda\mathbf{a}},\label{eq:involutionIsLinear}
\end{equation}
it is also continuous\footnote{Note that Cauchy's functional equation $\overline{\mathbf{a}\oplus\mathbf{b}}=\overline{\mathbf{a}}\oplus\overline{\mathbf{b}}$
for a continuous involution would allow us to deduce linearity. Here,
we can start by knowing the stronger property, linearity, which always
infers continuity over the reals.}.

Imposing a relatively strong choice, as is done here, risks constraining
allowable amplitude algebras to cases that no longer exhibit interesting
or useful structure. Given all possible permutations over a finite
set of arbitrary measurements and detectors, the percentage of symmetric
permutations generally decreases as the number of different detectors
grows. Correspondingly, the subspace of the amplitude algebra on which
we impose the real-number representation postulate becomes smaller
the more variety is allowed. Some follow-on constraints derive from
this, for example, for chains of symmetric paths to be represented
by the reals as well necessarily. However, these cases remain limited
in the overall scope of infinite variety and, therefore, for this
postulate to not overconstrain model representation.

\subsubsection{Continuous $p$-function}

With closure of the $p$-function,
\begin{equation}
p\left(\lambda\mathbf{a}\right)=p\left(\lambda\mathbf{1}\odot\mathbf{a}\right)=p\left(\lambda\mathbf{1}\right)p\left(\mathbf{a}\right),
\end{equation}
there exist functions $f:\mathbb{R}\rightarrow\mathbb{R}$ which only
depend on $\lambda,\mu\in\mathbb{R}$:
\begin{align}
f\left(\lambda\right) & :=p\left(\lambda\mathbf{1}\right)=\frac{p\left(\lambda\mathbf{1}\odot\mathbf{a}\right)}{p\left(\mathbf{a}\right)},\\
f\left(\lambda\mu\right) & =f\left(\lambda\right)f\left(\mu\right).
\end{align}
For continuous $f$, the general nonconstant solutions of this functional
equation\footnote{The constant solutions $f_{0}\left(\lambda\right)=0$ and $f_{1}\left(\lambda\right)=1$
do not give any information on probabilities of an experiment and
are not useful here.} to model a positive-valued $p$-function
given some $\alpha\in\mathbb{R}$ are of the type
\begin{equation}
f_{\alpha}\left(\lambda\right)=\left|\lambda\right|^{\alpha}.
\end{equation}

Going forward, we will assume $\alpha\neq0$ and continuity of these
$f$-functions, which makes the $p$-functions continuous as well:
\begin{equation}
p\left(\lambda\mathbf{a}\right)=\left|\lambda\right|^{\alpha}p\left(\mathbf{a}\right).\label{eq:pFunctionIsHomogeneousOfDegreeAlpha}
\end{equation}
Functions of this type are called \emph{positively homogeneous} of
degree $\alpha$.

We find for the edge cases
\begin{align}
p\left(\lambda\mathbf{1}\right)=0 & \quad\longrightarrow\quad\lambda=0,\label{eq:pLambda1eq0infersLambda0}\\
p\left(\lambda\mathbf{1}\right)=1 & \quad\longrightarrow\quad\lambda=\pm1.\label{eq:pLambda1eq1infersLambda1}
\end{align}

\subsection{\protect\label{subsec:Scalar-field}Scalar field}

Real numbers $\mathbb{R}$ are an algebraic field. To prove consistency
with the scalar representation of symmetric paths, we now trace all
field axioms for $\left\langle \mathbb{R},\oplus,\odot\right\rangle $
from the model where possible and otherwise show that they are not
in conflict:
\begin{itemize}
\item Coarsening of two symmetric paths always results in another symmetric
path, and the result is, therefore, again a scalar. This also applies
to chains of symmetric paths. The operation $\oplus$ is closed.
\item Additive associativity and commutativity of $\oplus$ are satisfied
from general coarsening, and additive inverses always exist.
\item Impossible paths have amplitude $\mathbf{0}$, which is the additive
identity, $\mathbf{a}\oplus\mathbf{0}=\mathbf{0}\oplus\mathbf{a}=\mathbf{a}$.
\item Chains of symmetric paths are also represented as scalars, and the
operation $\odot$ is closed.
\item Multiplicative associativity is satisfied from associativity of chaining
$\odot$.
\item Multiplicative commutativity is required by postulate from real-number
representation. It constrains the algebra but is not in conflict with
the model.
\item Paths that are trivial in their nonredundant form are also trivially
symmetric. Chaining with a trivial path is redundant and corresponds
to applying a scalar multiplicative unit, $1\mathbf{a}\equiv\mathbf{a}$.
\item Multiplicative inverses of $\mathbf{a}$ exist for $\mathbf{a}\neq\mathbf{0}$
as long as $p\left(\mathbf{a}\right)\neq0$. Because $p\left(\lambda\mathbf{1}\right)=0$
can only be true when $\lambda=0$, inverses exist for all nonzero
scalars.
\item Distributivity of multiplication over addition corresponds to distributivity
of chaining over coarsening in general.
\end{itemize}
With this, amplitudes representing chains of palindromic sequences
satisfy all field axioms when identifying field multiplication with
chaining and field addition with coarsening. We point out that most
of the field axioms trace from model properties of symmetric paths,
except for commutativity of multiplication, which is postulated.

Choosing the real numbers rules out other possible realizations of
an algebraic field. For example, requiring scalars to be ordered and
one-dimensional rules out the two-dimensional complex numbers $\mathbb{C}$
and specific field extensions (e.g. by supplying the rationals $\mathbb{Q}$
with a second dimension to basis $\sqrt{2}$). In one dimension, fixing
the reals excludes fields over a set with smaller cardinality (e.g.,
the rationals $\mathbb{Q}$), fields with characteristic not $0$
(e.g., finite fields, or the so-called ``nimbers'' which are infinite),
and fields over a set that is congruent to the reals but with a different
valuation (such as the $p$-adic numbers). These interesting opportunities
are advertised for exploration in follow-on work.

\subsection{Vector space}

We now recognize the amplitude space $\left\langle \mathbb{A},\oplus,\odot\right\rangle $
to be a vector space. In the most general case, a vector space $V$
over a field $K$ is a set with two operations,\emph{ }vector addition
$\oplus:V\times V\rightarrow V$ and scalar multiplication with a
vector $\otimes:K\times V\rightarrow V$, such that for all vectors
$\mathbf{a},\mathbf{b},\mathbf{c}\in V$ and scalars $\lambda,\mu\in K$,
there is
\begin{enumerate}
\item $\left(\mathbf{a}\oplus\mathbf{b}\right)\oplus\mathbf{c}=\mathbf{a}\oplus\left(\mathbf{b}\oplus\mathbf{c}\right)$,
\item $\mathbf{a}\oplus\mathbf{b}=\mathbf{b}\oplus\mathbf{a}$,
\item there exists an element $\mathbf{0}\in V$ such that $\mathbf{0}\oplus\mathbf{a}=\mathbf{a}\oplus\mathbf{0}=\mathbf{a}$,
\item there exists an element $\left(-\mathbf{a}\right)\in V$ such that
$\mathbf{a}\oplus\left(-\mathbf{a}\right)=\mathbf{0}$,
\item $\lambda\otimes\left(\mu\otimes\mathbf{a}\right)=\left(\lambda\mu\right)\otimes\mathbf{a}$,
\item $1\otimes\mathbf{a}=\mathbf{a}$,
\item $\lambda\otimes\left(\mathbf{a}\oplus\mathbf{b}\right)=\lambda\otimes\mathbf{a}\oplus\lambda\otimes\mathbf{b}$,
\item $\left(\lambda+\mu\right)\otimes\mathbf{a}=\lambda\otimes\mathbf{a}\oplus\mu\otimes\mathbf{a}$.
\end{enumerate}
From these general axioms, we identify elements $\mathbf{a},\mathbf{b},\mathbf{c}\in V$
with amplitudes $\mathbf{a},\mathbf{b},\mathbf{c}\in\mathbb{A}$,
the field $K$ with the reals $\mathbb{R}$, and scalars $\lambda,\mu\in K$
with amplitudes $\lambda\mathbf{1},\mu\mathbf{1}\in\mathbb{A}$. Axioms
``1.'' and ``2.'' then follow directly from coarsening. The additive
identity $\mathbf{0}\in\mathbb{A}$ represents impossible paths (``3.''),
and amplitude addition is invertible (``4.''). Scalars $\lambda\equiv\lambda\mathbf{1}$
mix with general amplitudes exactly as in ``5.'' through ``8.''
when identifying scalar multiplication with amplitude multiplication,
$\otimes\equiv\odot$.

There are known theorems for faithful representation of a vector space.
A linearly independent set of vectors that spans the entire vector
space is called a basis of the vector space. A vector space generally
has many bases, all of which have the same cardinality. This is the
dimension of the vector space.

After fixing such a basis $\mathbf{e}_{V}$ of an $n$-dimensional
vector space $V$ to the $n$-tuple $\mathbf{e}_{V}:=\left(\mathbf{e}_{0},\ldots,\mathbf{e}_{n-1}\right)$,
$\mathbf{e}_{r}\in V$ for $r=0,\ldots,n-1$, every vector $\mathbf{a}\in V$
can be written uniquely as a linear combination of these basis elements
with scalars $a_{r}\in K$:
\begin{equation}
\mathbf{a}=:\sum_{r=0}^{n-1}a_{r}\mathbf{e}_{r}.
\end{equation}
The scalars $a_{r}$ are called coordinates (or coefficients) of the
vector $\mathbf{a}$, which can then be represented as tuple to the
corresponding basis elements in $\mathbf{e}_{V}$:
\begin{equation}
\mathbf{a}=\left(a_{0},\ldots,a_{n-1}\right).
\end{equation}

\subsection{Algebra over a field}

The configuration space of amplitudes is a real vector space that
is further equipped with a bilinear product, i.e., a multiplication
$\odot$ that is both homogeneous
\begin{equation}
\left(\lambda\mathbf{a}\right)\odot\left(\mu\mathbf{b}\right)=\lambda\mu\left(\mathbf{a}\odot\mathbf{b}\right),
\end{equation}
and distributes over addition $\oplus$,
\begin{align}
\mathbf{a}\odot\left(\mathbf{b}\oplus\mathbf{c}\right) & =\mathbf{a}\odot\mathbf{b}\oplus\mathbf{a}\odot\mathbf{c},\\
\left(\mathbf{b}\oplus\mathbf{c}\right)\odot\mathbf{a} & =\mathbf{b}\odot\mathbf{a}\oplus\mathbf{c}\odot\mathbf{a}.
\end{align}
This characterizes the amplitude algebra $\mathbb{A}$ as a conventional
algebra over a field, here over the reals $\mathbb{R}$.

Given a basis $\mathbf{e}_{V}:=\left(\mathbf{e}_{0},\ldots,\mathbf{e}_{n-1}\right)$
of an $n$-dimensional vector space $V$, structure constants $f_{ijk}\in K$
with $i,j,k=0,\ldots,n-1$ allow to explicitly represent the product
of $\mathbb{A}$, by specifying the product of every pair of basis
vectors $\mathbf{e}_{i},\mathbf{e}_{j}$ as a linear combination:
\begin{equation}
\mathbf{e}_{i}\odot\mathbf{e}_{j}:=\bigoplus_{k=0}^{n-1}f_{ijk}\mathbf{e}_{k}.\label{eq:StruconsProduct}
\end{equation}
Because every vector $\mathbf{a}\in V$ can be written as a linear
combination of the $\mathbf{e}_{i}$, this product extends to all
vectors in the vector space.

Every product from an algebra over a field can be uniquely represented
by structure constants (after fixing a basis). Conversely, any product
represented by structure constants is from an algebra over a field.

\subsection{$p$-functions as signomials}

The positively homogeneous $p$-functions of degree $\alpha$ over
vectors $\mathbf{a}=\left(a_{0},\ldots,a_{n-1}\right)$ are now expressed
as generalized polynomials over the vector coefficients. Referred
to as a signomials, these are sums of arbitrarily many (generalized)
monomials, indexed $s=1,2,3,\ldots$ and weighted by some $c_{s}\in\mathbb{R}$.
These monomials are products made from the coefficients of $\mathbf{a}$,
forced nonnegative, and taken to some real powers $\beta\left(s,r\right)\in\mathbb{R}_{\geq0}$
that add up to $\alpha$:
\begin{equation}
p\left(\mathbf{a}\right):=\sum_{\forall s}c_{s}\left(\prod_{r=0}^{n-1}\left|a_{r}\right|^{\beta\left(s,r\right)}\right),\textrm{ with }\sum_{r=0}^{n-1}\beta\left(s,r\right)=\alpha\textrm{ for all }s\label{eq:pAsHomogeneousSignomial}
\end{equation}
(using $0^{0}:=1$ by convention). Since the exponents in all monomials
add up to $\alpha$, this can be termed a \emph{homogeneous signomial}\footnote{For contemporary use of term, and signomials in general, see e.g.
\cite{DM2022}.} of degree $\alpha$.

The reader can readily verify that (\ref{eq:pAsHomogeneousSignomial})
solves the homogeneity equation for the $p$-functions\footnote{The authors were not able to find completeness or uniqueness theorems
for signomials towards solving the homogeneity equation in the current
literature, or likewise, for solutions of the multiplicative Cauchy
equation (Cauchy's ``power equation'') $p\left(\mathbf{x}\odot\mathbf{y}\right)=p\left(\mathbf{x}\right)p\left(\mathbf{y}\right)$
for general vector products $\mathbf{x}\odot\mathbf{y}$. Works like
\cite{AD1989} give solutions for matrix products in the most general
case. It may be interesting to ask whether, for example, the expression
$\left(a_{0}^{2}+a_{1}^{2}\right)^{\pi}$ for $a_{0},a_{1}\in\mathbb{R}$
can be written as an infinite sum of products of the $a_{0},a_{1}$
taken to exponents that add up to $2\pi$, i.e., $\left(a_{0}^{2}+a_{1}^{2}\right)^{\pi}=\left|a_{0}\right|^{2\pi}+\left|a_{1}\right|^{2\pi}+\ldots+c_{s}\left|a_{0}\right|^{\beta\left(s,0\right)}\left|a_{1}\right|^{\beta\left(s,1\right)}+\ldots$
with $\beta\left(s,0\right)+\beta\left(s,1\right)=2\pi$ for all $s$.
Since signomials of the form (\ref{eq:pAsHomogeneousSignomial}) will
prove to be sufficient to specify the $p$-function in this paper,
we leave these follow-on questions for the interested reader.}, $p\left(\lambda\mathbf{a}\right)=\left|\lambda\right|^{\alpha}p\left(\mathbf{a}\right)$.

\section{\protect\label{sec:Born-rule}Born rule}

Connecting to canonical physics, we now show that $p$-function is
quadratic in amplitudes, and interpret it as a generalized Born rule
from Quantum Mechanics. Using well-known axioms from mathematics,
we constrain allowable amplitude algebras to the small, distinct set
of associative composition algebras over the reals. This includes
the complex numbers, concluding our argument towards the suitability
of the model and methodology in this paper.

\subsection{Quadratic form $Q\left(\mathbf{a}\right)$}

The product $\mathbf{a}\odot\overline{\mathbf{a}}$ is self-adjoint
for any $\mathbf{a}\in\mathbb{A}$, and therefore, always corresponds
to a scalar, $\mathbf{a}\odot\overline{\mathbf{a}}=\lambda\mathbf{1}$,
$\lambda\in\mathbb{R}$, per equation (\ref{eq:selfAdjointAmplIsAlwaysScalar}).
We now define a map $Q:\mathbb{A}\rightarrow\mathbb{R}$ as:
\begin{align}
Q\left(\mathbf{a}\right)\mathbf{1} & :=\mathbf{a}\odot\overline{\mathbf{a}}=\lambda\mathbf{1}.\label{eq:defQuadraticForm}
\end{align}
Given $\mu\in\mathbb{R}$, there is:
\[
Q\left(\mu\mathbf{a}\right)\mathbf{1}=\left(\mu\mathbf{1}\odot\mathbf{a}\right)\odot\overline{\left(\mu\mathbf{1}\odot\mathbf{a}\right)}=\mu^{2}Q\left(\mathbf{a}\right)\mathbf{1}.
\]
Because $Q$ is positively homogeneous of degree $2$, it can be expressed
as a homogeneous signomial over the coefficients of $\mathbf{a}$
similar to the $p$-functions. Furthermore, since it is a product
from an algebra over a field, expressible through summation of structure
constants, it is, in fact, a homogeneous polynomial of degree $2$.
This makes $Q$ a quadratic form.

From associativity of amplitudes and commutativity of scalars follows
\begin{align}
Q\left(\mathbf{a}\right)\mathbf{1}\odot\mathbf{a} & =\left(\mathbf{a}\odot\overline{\mathbf{a}}\right)\odot\mathbf{a}=\mathbf{a}\odot\left(\overline{\mathbf{a}}\odot\mathbf{a}\right)=\mathbf{a}\odot Q\left(\overline{\mathbf{a}}\right)\mathbf{1}=Q\left(\overline{\mathbf{a}}\right)\mathbf{1}\odot\mathbf{a},\\
Q\left(\mathbf{a}\right) & =Q\left(\overline{\mathbf{a}}\right).\label{eq:QisQofInvolution}
\end{align}
Also, $Q$ is multiplicative since for any $\mathbf{b}\in\mathbb{A}$
with $\mathbf{b}\odot\overline{\mathbf{b}}=\mu\mathbf{1}$, $\mu\in\mathbb{R}$
there is
\begin{align}
Q\left(\mathbf{a}\odot\mathbf{b}\right)\mathbf{1} & =\mathbf{a}\odot\mathbf{b}\odot\overline{\mathbf{a}\odot\mathbf{b}}=\mathbf{a}\odot\mathbf{b}\odot\overline{\mathbf{b}}\odot\overline{\mathbf{a}}\nonumber \\
 & =\mathbf{a}\odot\left(\mathbf{b}\odot\overline{\mathbf{b}}\right)\odot\overline{\mathbf{a}}=\mu\mathbf{a}\odot\overline{\mathbf{a}}=\lambda\mu\mathbf{1}\label{eq:quadraticFormHasProductCompositionProperty}\\
 & =Q\left(\mathbf{a}\right)Q\left(\mathbf{b}\right)\mathbf{1}.\nonumber 
\end{align}

\subsection{Quadratic probabilities}

To show the behavior of the $p$-functions under scaling, we first
look at a symmetric experiment $\,^{\circledcirc}\negthinspace\mathcal{M}=\left[\mathbf{N},\mathbf{M},\mathbf{N}\right]$,
where measurement $\mathbf{N}$ has at least one atomic detector $N$,
and $\mathbf{M}$ has $s$ atomic detectors, $\mathbf{M}=\left\{ M^{1},M^{2},\ldots M^{s}\right\} =\left\{ \left\{ m^{1}\right\} ,\left\{ m^{2}\right\} ,\ldots\left\{ m^{s}\right\} \right\} $.
Fixing $N$ as source and target, the possible symmetric paths $\,^{\circledcirc}\negthinspace A^{r}\in\mathcal{S}_{^{\circledcirc}\negthinspace\mathcal{M}}$
with $r=1\ldots s$ are:
\begin{equation}
\,^{\circledcirc}\negthinspace A^{r}:=\left[N,M^{r},N\right].
\end{equation}
When coarse-graining over all paths, $^{\cup}A:=\bigvee_{r=1}^{s}\,^{\circledcirc}\negthinspace A^{r}=\left[N,\,^{\cup}M,N\right]$,
the measurement
\begin{equation}
^{\cup}\mathbf{M}=\left\{ ^{\cup}M\right\} =\left\{ \bigcup_{r=1}^{s}M^{r}\right\} =\left\{ \left\{ m^{1},m^{2},\ldots m^{s}\right\} \right\} 
\end{equation}
is certain and may be removed from $^{\cup}A$ without changing its
probability. The remaining trivial path $\left[N,N\right]$ has unit
amplitude, such that the fully coarse-grained path must have unit
probability,
\begin{equation}
P\left(^{\cup}A\right)=P\left(\bigvee_{r=1}^{s}\,^{\circledcirc}\negthinspace A^{r}\right)=1.
\end{equation}
All paths $\,^{\circledcirc}\negthinspace A^{r}$ are symmetric and
therefore represented by scalars, making $\phi\left(^{\cup}A\right)=\bigoplus_{r=1}^{s}\phi\left(\,^{\circledcirc}\negthinspace A^{r}\right)=\lambda\mathbf{1}$
a scalar as well. $P\left(^{\cup}A\right)=p\left(\phi\left(^{\cup}A\right)\right)=p\left(\lambda\mathbf{1}\right)=1$
infers $\lambda=\pm1$ per (\ref{eq:pLambda1eq1infersLambda1}), and
fully coarsening results in the unit amplitude up to sign:
\begin{equation}
\phi\left(^{\cup}A\right)=\mathbf{\pm1}.\label{eq:fullyCoarseningScalarsIsUnitAmplitude}
\end{equation}

Paths $\,^{\circledcirc}\negthinspace A^{r}$ can be factorized into
pairs of adjoint paths because all detectors $M^{r}:=\left\{ m^{r}\right\} $
are atomic:
\begin{align}
\,^{\circledcirc}\negthinspace A^{r} & \sim\left[N,M^{r}\right]\cdot\left[M^{r},N\right].
\end{align}
With amplitudes $\mathbf{a}^{r}:=\phi\left(\left[N,M^{r}\right]\right)$
we have therefore:
\begin{align}
\phi\left(A^{r}\right) & =\phi\left(\left[N,M^{r}\right]\right)\odot\phi\left(\left[M^{r},N\right]\right)=\mathbf{a}^{r}\odot\overline{\mathbf{a}^{r}},\\
\phi\left(^{\cup}A\right) & =\bigoplus_{r=1}^{s}\left(\mathbf{a}^{r}\odot\overline{\mathbf{a}^{r}}\right)=\mathbf{\pm1}.
\end{align}
Using the quadratic form $Q$, this becomes:
\begin{align}
\bigoplus_{r=1}^{s}\left(\mathbf{a}^{r}\odot\overline{\mathbf{a}^{r}}\right) & =\bigoplus_{r=1}^{s}Q\left(\mathbf{a}^{r}\right)\mathbf{1}=\mathbf{\pm1},\\
\sum_{r=1}^{s}Q\left(\mathbf{a}^{r}\right) & =\pm1.\label{eq:fullyCoarsenedSumQisOne}
\end{align}

Conversely, the probability of independently observing any of the
paths $\left[N,M^{r}\right]$ must also be $1$ per section \ref{subsec:Connection-to-classical},
\begin{equation}
\sum_{r=1}^{s}P\left(\left[N,M^{r}\right]\right)=\sum_{r=1}^{s}p\left(\mathbf{a}^{r}\right)=1.\label{eq:fullyCoarsenedSumPisOne}
\end{equation}

The sum of quadratic forms as in (\ref{eq:fullyCoarsenedSumQisOne})
is again a quadratic form, and the space of signomials of the same
degree is closed under summation likewise (\ref{eq:fullyCoarsenedSumPisOne}).
For both these sums to always add up to $+1$ over the same coefficients
from all $\mathbf{a}_{r}$, the exponential behavior of all summands
from the argument components must be identical\footnote{Because the
degree of $Q$ is known to be 2, this follows from the identity theorem
for polynomials for positive integral $\alpha$. Defining $p$ as a
signomial -- or any other function over $\mathbb{R}^n$ -- does not alter this
conclusion, given the identity of the functions over an open set in $\mathbb{R}^n$.}.
The other case of all $Q$ adding up to $-1$ allows the same conclusion
since $-Q\left(\nu\mathbf{a}\right) = \nu^{2} \left(- Q\left(\mathbf{a}\right)\right)$,
which makes the sign in $\pm Q$ a one-time choice for the experiment. Chosing
the positive sign by convention this identifies $\alpha=2$
and the $p$-functions with $Q$,
\begin{align}
p & \equiv Q,\\
p\left(\nu\mathbf{1}\right) & =\nu^{2},\\
p\left(\nu\mathbf{a}\right) & =\nu^{2}p\left(\mathbf{a}\right).
\end{align}
Because the $p$-functions model probability, when traced from valid
path operations in the model, we have effectively shown that probabilities
are quadratic in transition amplitudes between measurements. This
can be interpreted as a generalized Born rule, in the sense of Feynman's
rules as in \cite{GKS2010}.

\subsection{$p$-functions are a nondegenerate quadratic form}

To fully utilize our configuration space of the amplitude algebra,
we want the $p\left(\mathbf{a}\right)$ to be nondegenerate: They
must require $n$ independent parameters in the $n$-dimensional vector
space of $\mathbb{A}$ and cannot be reduced to a function over a
lower-dimensional subspace of $\mathbb{A}$.

This can be formulated in a coordinate-independent way by using the
associated bilinear form $B:\mathbb{A}\times\mathbb{A}\rightarrow\mathbb{R}$,
\begin{equation}
B\left(\mathbf{a},\mathbf{b}\right)\mathbf{1}:=\left(p\left(\mathbf{a}\oplus\mathbf{b}\right)-p\left(\mathbf{a}\right)-p\left(\mathbf{b}\right)\right)\mathbf{1}=\mathbf{a}\odot\overline{\mathbf{b}}\oplus\mathbf{b}\odot\overline{\mathbf{a}}.\label{eq:defBilinearForm}
\end{equation}
Degeneracy of $B$ and, by extension, $p$, then corresponds to existence
of a nonzero $\mathbf{b}_{\bot}\in\mathbb{A}$ for which $B\left(\mathbf{a},\mathbf{b}_{\bot}\right)=0$
for all $\mathbf{a}$. Such a nondegenerate $B$ allows to conclude
for any given $\mathbf{c}\in\mathbb{A}$
\begin{equation}
B\left(\mathbf{a},\mathbf{c}\right)=0\textrm{ for all }\mathbf{a}\in\mathbb{A}\rightarrow\mathbf{c}=\mathbf{0}.
\end{equation}

To show equivalence of such nondegeneracy with our requirement for
$n$ linear independent arguments modeling $p$ in an $n$-dimensional
vector space, we show that a degenerate $p$ can always be modeled
with fewer than $n$ independent parameters; and conversely that a
nondegenerate $p$ requires $n$ parameters.

First, we assume that $p$ would be degenerate, and write $\mathbf{b}_{\bot}\in\mathbb{A}$
for a nonzero amplitude that has $B\left(\mathbf{a},\mathbf{b}_{\bot}\right)=0$
for all $\mathbf{a}$. In that case equation (\ref{eq:defBilinearForm})
becomes $p\left(\mathbf{a}\oplus\mathbf{b}_{\bot}\right)=p\left(\mathbf{a}\right)+p\left(\mathbf{b}_{\bot}\right)$
and $p$ is an additive map for every $\mathbf{a}$. This is trivially
satisfied if $p$ is constant $0$. For nonconstant $p$, Cauchy's
functional equation would require the continuous $p$ to be linear
in $\mathbb{R}$. Since we already established $p$ to be quadratic
in general, this can only be the case for general $\mathbf{a}$ if
$p\left(\mathbf{b}_{\bot}\right)=0$. This makes $p\left(\mathbf{a}\oplus\mu\mathbf{b}_{\bot}\right)=p\left(\mathbf{a}\right)+\mu^{2}p\left(\mathbf{b}_{\bot}\right)=p\left(\mathbf{a}\right)$
for any $\mu\in\mathbb{R}$, and $p$ cannot depend on the one-dimensional
space spanned by $\mu\mathbf{b}_{\bot}$. That is, degenerate $p$
can always be reduced to a function that excludes the parameters where
$p\left(\mu\mathbf{b}_{\bot}\right)=0$ and, therefore, fully understood
using fewer than $n$ independent parameters.

Conversely, now assuming $p$ would be nondegenerate but depending
on fewer than $n$ independent coefficients of its argument. It would
mean that a coordinate basis exists in which the $0$-coefficient
of any vector does not contribute to the value of $p$. In such a
basis, choose a vector $\mathbf{b}^{\prime}:=\left(b_{0}^{\prime},0,\ldots,0\right)$
that differs from the zero-vector $\left(0,\ldots,0\right)$ only
by $b_{0}^{\prime}\neq0$. Since $p\left(\mathbf{b}^{\prime}\right)$
does not depend on the $0$-component of its argument, there is always
$p\left(\mathbf{b}^{\prime}\right)=p\left(\mathbf{0}\right)=0$, and
likewise $p\left(\mathbf{a}\oplus\mathbf{b}^{\prime}\right)=p\left(\mathbf{a}\right)$.
With this, $B\left(\mathbf{a},\mathbf{b}^{\prime}\right)=p\left(\mathbf{a}\oplus\mathbf{b}^{\prime}\right)-p\left(\mathbf{a}\right)-p\left(\mathbf{b}^{\prime}\right)=0$
for every $\mathbf{a}$, which contradicts the requirement for $p$
being nondegenerate.

\subsection{Composition algebras}

Per \cite{Jacobson1958}, a unital algebra $\mathbb{A}$ over the
reals\footnote{The treatment in \cite{Jacobson1958} is more general in that it allows
not just the real number field but any field $K$ of characteristic
not 2. We will not go into details here since we have already chosen
$K=\mathbb{R}$ per postulate.} is a \emph{composition algebra} if it admits a nondegenerate quadratic
form $Q:\mathbb{A}\rightarrow\mathbb{R}$ with product composition.
This means that there exists a multiplicative unit $\mathbf{1}\in\mathbb{A}$,
and for every $\mathbf{a},\mathbf{b}\in\mathbb{A}$, $\lambda\in\mathbb{R}$
there is:
\begin{align}
Q\left(\mathbf{a}\odot\mathbf{b}\right) & =Q\left(\mathbf{a}\right)Q\left(\mathbf{b}\right),\\
Q\left(\lambda\mathbf{a}\right) & =\lambda^{2}Q\left(\mathbf{a}\right),\\
B\left(\mathbf{a},\mathbf{b}\right) & :=Q\left(\mathbf{a}+\mathbf{b}\right)-Q\left(\mathbf{a}\right)-Q\left(\mathbf{b}\right),\\
B\left(\mathbf{a},\mathbf{c}\right)=0\textrm{ for all }\mathbf{a}\in\mathbb{A} & \rightarrow\mathbf{c}=\mathbf{0}.
\end{align}
We have traced all these axioms from the model, concluding that any
allowable amplitude algebra must be a composition algebra.

Only a few distinct algebras over the real numbers admit these properties.
Next to the field algebra $\mathbb{R}$ itself are the complex $\mathbb{C}$,
quaternion $\mathbb{H}$, and octonion $\mathbb{O}$ algebras over
vector spaces $\mathbb{R}^{2}$, $\mathbb{R}^{4}$, and $\mathbb{R}^{8}$,
respectively, with their split forms $\mathbb{C}^{\prime}$, $\mathbb{H}^{\prime}$,
and $\mathbb{O}^{\prime}$. Because we require multiplication to be
associative, the nonassociative octonions $\mathbb{O}$ and split-octonions
$\mathbb{O}^{\prime}$ are ruled out here\footnote{For composition algebras in general, and these possibilities over
the reals in particular, see, for example \cite{PM:CompAlg,WP:CompAlg}.}.

Our $p$-function on amplitudes is the well-known quadratic form $Q$
on these algebras, which may either be positive-definite (for $\mathbb{R}$,
$\mathbb{C}$, $\mathbb{H}$) or have split-signature (for $\mathbb{C}^{\prime}$,
$\mathbb{H}^{\prime}$). For $\mathbb{R}$, $\mathbb{C}$ and $\mathbb{H}$
it is the Euclidean norm of $\mathbf{a}$ in one, two, and four dimensions,
respectively:
\begin{align}
p_{\mathbb{R}}\left(\mathbf{a}\right)=p_{\mathbb{R}}\left(\left(a_{0}\right)\right) & =a_{0}^{2},\\
p_{\mathbb{C}}\left(\mathbf{a}\right)=p_{\mathbb{C}}\left(\left(a_{0},a_{1}\right)\right) & =a_{0}^{2}+a_{1}^{2},\\
p_{\mathbb{H}}\left(\mathbf{a}\right)=p_{\mathbb{H}}\left(\left(a_{0},a_{1},a_{2},a_{3}\right)\right) & =a_{0}^{2}+a_{1}^{2}+a_{2}^{2}+a_{3}^{2}.
\end{align}
Conversely, for the split-algebras $\mathbb{C}^{\prime}$ and $\mathbb{H}^{\prime}$
it is an isotropic quadratic form that is hyperbolic\footnote{Given a suitable labeling of basis elements. We will not further go
into the details of split-algebras in this paper, but point to popular
introductions, e.g., as in \cite{WP:SplitQuaternion}.}:
\begin{align}
p_{\mathbb{C}^{\prime}}\left(\mathbf{a}\right)=p_{\mathbb{C}^{\prime}}\left(\left(a_{0},a_{1}\right)\right) & =a_{0}^{2}-a_{1}^{2},\\
p_{\mathbb{H}^{\prime}}\left(\mathbf{a}\right)=p_{\mathbb{H}^{\prime}}\left(\left(a_{0},a_{1},a_{2},a_{3}\right)\right) & =\left(a_{0}^{2}+a_{1}^{2}\right)-\left(a_{2}^{2}+a_{3}^{2}\right).
\end{align}

We point out that the dimensionality of allowable amplitude algebras,
$\dim\mathbb{A}\in\left\{ 1,2,4\right\} $, has not been prescribed
or postulated by us anywhere but follows solely from the nature of
composition algebras.

Expressions in conventional Quantum Mechanics typically are over complex
numbers and often use matrix formulations. While quaternions or split-algebras
may not appear familiar at first sight, they are effectively already
part of the standard toolset of a physicist: The familiar Pauli matrices
are typically represented as $2\times2$ complex matrices and can
be used to construct the imaginary parts of a quaternion basis. The
algebra of $2\times2$ real matrices is, in fact, a split-quaternion
algebra\cite{WP:SplitQuaternion}.

With this, we have demonstrated the utility of complex numbers in
our model and related composition algebras already used in canonical
physics today. This supports our claim that the model can be a building
block towards a faithful, complete reconstruction of the laws of Quantum
Mechanics.

\subsection{Alternative axioms towards composition algebras}

Next to the axioms used above, Nathan Jacobson gave two more alternative
sets of axioms, each leading to the same composition algebras over
the reals. In order to strengthen our argument for internal consistency
of our axioms and choices, we now discuss these as well\footnote{While the power of axiomatic, deductive reasoning could be paraphrased
as ``proven once, valid everywhere,'' choice representations may
carry doubt of their universality. This could be put pointedly maybe
as ``chosen once, to be proven everywhere.'' In practice, the situation
is not nearly as dire, as choices at some point become axioms by their
very nature. Nevertheless, until a complete axiomatic description
of Quantum Mechanics is posed, it is prudent to check for internal
consistency between our current choices and the more foundational
axioms.}.

Also in \cite{Jacobson1958}, a unital alternative algebra $\mathbb{A}$
over $\mathbb{R}$ is a composition algebra if it admits an involution
$\mathbf{a}\rightarrow\overline{\mathbf{a}}$, such that both product
and sum of any $\mathbf{a}\in\mathbb{A}$ with its involution $\overline{\mathbf{a}}$
is a real form. This means that there exists a multiplicative unit
$\mathbf{1}\in\mathbb{A}$, and for every $\mathbf{a},\mathbf{b}\in\mathbb{A}$,
there is:
\begin{align}
\mathbf{a}\odot\left(\mathbf{a}\odot\mathbf{b}\right) & =\left(\mathbf{a}\odot\mathbf{a}\right)\odot\mathbf{b},\\
\mathbf{b}\odot\left(\mathbf{a}\odot\mathbf{a}\right) & =\left(\mathbf{b}\odot\mathbf{a}\right)\odot\mathbf{a},\\
\overline{\overline{\mathbf{a}}} & =\mathbf{a},\\
\mathbf{a}\odot\overline{\mathbf{a}} & =Q\left(\mathbf{a}\right)\mathbf{1}\textrm{ with }Q:\mathbb{A}\rightarrow\mathbb{R},\\
\mathbf{a}\oplus\overline{\mathbf{a}} & =T\left(\mathbf{a}\right)\mathbf{1}\textrm{ with }T:\mathbb{A}\rightarrow\mathbb{R}.
\end{align}
We readily recognize all axioms but the last one from our work here.
Interestingly, these axioms do not mention the product composition
property of the $Q$ or existence of a quadratic form. Instead, properties
of the involution are required, which in turn are absent from the
earlier axiom set.

Algebraically, we can immediately verify that the $T$-function (the
``trace'') is indeed real from (\ref{eq:defBilinearForm}):
\begin{equation}
B\left(\mathbf{a},\mathbf{1}\right)\mathbf{1}=\mathbf{a}\oplus\overline{\mathbf{a}}.
\end{equation}
This is, therefore, consistent with our model. However, this amplitude
expression can never represent an allowable path operation,
\begin{equation}
\phi\left(A\overset{??}{\vee}\overline{A}\right)=\mathbf{a}\oplus\overline{\mathbf{a}}.
\end{equation}
Looking at $A\overset{??}{\vee}\overline{A}$ by itself, $A$ would
have to differ from $\overline{A}$ by exactly one measurement result
to permit coarsening, which is impossible. Conversely, in an expression
\begin{equation}
A\overset{??}{\vee}\left(\overline{A}\vee B\right)
\end{equation}
with some path $B$, the cardinality of one of the measurement results
in $A$ would have to be larger than the corresponding one in $\overline{A}$,
which is a contradiction. The trace function is an example of an algebraic
property of the representation space that is formally always true
but can never be traced from the model; and is, therefore, not in
conflict with the model, either.

We can however trace a third set of axioms towards composition algebras,
given in \cite{Jacobson1981}. Only looking at the real algebra case,
an alternative algebra $\mathbb{A}$ over $\mathbb{R}$ is a composition
algebra if it is free from absolute zero divisors and admits an involution
$\mathbf{a}\rightarrow\overline{\mathbf{a}}$ that is an anti-isomorphism,
where the product of any $\mathbf{a}\in\mathbb{A}$ with its involution
$\overline{\mathbf{a}}$ is a real form. This means that for every
$\mathbf{a},\mathbf{b},\mathbf{c}\in\mathbb{A}$ there is:
\begin{align}
\mathbf{a}\odot\left(\mathbf{a}\odot\mathbf{b}\right) & =\left(\mathbf{a}\odot\mathbf{a}\right)\odot\mathbf{b},\\
\mathbf{b}\odot\left(\mathbf{a}\odot\mathbf{a}\right) & =\left(\mathbf{b}\odot\mathbf{a}\right)\odot\mathbf{a},\\
\overline{\overline{\mathbf{a}}} & =\mathbf{a},\\
\mathbf{a}\odot\overline{\mathbf{a}} & =Q\left(\mathbf{a}\right)\mathbf{1}\textrm{ with }Q:\mathbb{A}\rightarrow\mathbb{R},\\
\overline{\mathbf{a}\odot\mathbf{b}} & =\overline{\mathbf{b}}\odot\overline{\mathbf{a}},\\
\left(\mathbf{c}\odot\mathbf{a}\right)\odot\mathbf{c}=\mathbf{0}\textrm{ with }\mathbf{a}\neq\mathbf{0} & \rightarrow\mathbf{c}=\mathbf{0}.
\end{align}
Again, all but the last of the axioms are familiar and were previously
shown to trace from the model. As for the last item, for $\left(\mathbf{c}\odot\mathbf{a}\right)$
to result in a zero amplitude for any nonzero $\mathbf{a}$, the $\mathbf{c}$
must represent an impossible path and be $\mathbf{0}$, making the
entire expression $\left(\mathbf{c}\odot\mathbf{a}\right)\odot\mathbf{c}=\mathbf{0}$.
Where $\left(\mathbf{c}\odot\mathbf{a}\right)$ would be nonzero,
conversely, chaining another $\odot\mathbf{c}$ could make that expression
$\mathbf{0}$ only if $\mathbf{c}=\mathbf{0}$, which is a contradiction.
Amplitude algebra is, therefore, free from absolute zero divisors,
and all axioms are traced from our model.

\subsection{A note on split-algebras}

Split-complex numbers $\mathbb{C}^{\prime}$ and split-quaternions
$\mathbb{H}^{\prime}$ contain subspaces where the $p$-function is
$0$ for some nonzero amplitudes. In $\mathbb{C}^{\prime}$ with $p_{\mathbb{C}^{\prime}}\left(\mathbf{a}\right)=a_{0}^{2}-a_{1}^{2}$,
these are the nonzero diagonals in the plane, $a_{0}=\pm a_{1}$.
In $\mathbb{H}^{\prime}$ with $p_{\mathbb{H}^{\prime}}\left(\mathbf{a}\right)=\left(a_{0}^{2}+a_{1}^{2}\right)-\left(a_{2}^{2}+a_{3}^{2}\right)$
these are the hyperplanes with $a_{0}^{2}+a_{1}^{2}=a_{2}^{2}+a_{3}^{2}$.
As section \ref{subsec:ImpossiblePathsAsAdditiveIdentity} points
out, such algebraic subspaces must never represent allowable paths
in the model algebra. Furthermore, $\mathbb{C}^{\prime}$-spaces $\left|a_{0}\right|<\left|a_{1}\right|$
and $\mathbb{H}^{\prime}$-spaces $a_{0}^{2}+a_{1}^{2}<a_{2}^{2}+a_{3}^{2}$
have a negative $p$-function, which can never occur in $\mathbb{C}$
or $\mathbb{H}$.

While this may appear undesirable at first, the situation is essentially
no different, e.g., in the complex numbers $\mathbb{C}$ where numbers
outside the unit circle have a $p$-function value greater than $1$,
which would be nonsensical when interpreted as probability.

To be consistent with our model, we do have to satisfy one requirement:
As long as the $p$-function represents paths with nonzero probability,
multiplication must be invertible (per section \ref{subsec:MultiplicativeInverse}).
This is indeed the case: Given any $\mathbf{a}\in\mathbb{A}\in\left\{ \mathbb{R},\mathbb{C},\mathbb{C}^{\prime},\mathbb{H},\mathbb{H}^{\prime}\right\} $,
multiplicative inverses $\mathbf{a}^{-1}\in\mathbb{A}$ are well-defined
for nonzero $p\left(\mathbf{a}\right)$:
\begin{equation}
\mathbf{a}^{-1}=\frac{\overline{\mathbf{a}}}{p_{\mathbb{A}}\left(\mathbf{a}\right)}\quad\left(p_{\mathbb{A}}\left(\mathbf{a}\right)\neq0\right).
\end{equation}
This is, therefore, consistent with the required existence of multiplicative
inverses for amplitudes representing allowable paths.

Compared to the split-algebras with their hyperbolic $p$-function,
the circular geometry in $\mathbb{C}$ and $\mathbb{H}$ may appear
simpler or more symmetric when separating formally defined abstract
amplitudes from those with $p\in\left[0,1\right]$ that may represent
concrete paths in the model. However, this does not exclude split-algebras
from utility towards reconstructing Quantum Mechanics: The experiment
ultimately verifies nature's geometry, regardless of how simple or
symmetric we may find it.

\section{\protect\label{sec:Summary-and-outlook}Summary and outlook}

\subsection{What has and has not been accomplished}

In this paper, we explored a model that was first advertised
in \cite{GKS2010} as a building block in the Quantum Reconstruction
Program, which included reconstructing core algebraic properties brought
initially forward by Richard P.~Feynman. Addressing open questions
posed in \cite{Koepl2023gks}, we confirmed that it is possible to
represent probabilities from the model by amplitudes on complex numbers.
Going beyond the original work, we did so coordinate-independently
without requiring a two-dimensional configuration space a priori.
Furthermore, we traced algebraic unit elements, scalar multiplication,
and most field axioms for scalars from the model. Demonstrating that
scalars form an algebraic field, including invertibility of addition
and multiplication, allowed us to conclude that the configuration
space of amplitudes is indeed a vector space\footnote{We believe this to be a significant innovation, also with thanks to
John A.~Shuster for pointing out a counterexample early on: If one
were to define multiplication $\odot$ and addition $\oplus$ between
a pair of two-dimensional vectors $\mathbf{a}$ and $\mathbf{b}$
in terms of familiar complex number operations, $\mathbf{a}\odot\mathbf{b}:=\exp\left(\ln\mathbf{a}\ln\mathbf{b}\right)$,
$\mathbf{a}\oplus\mathbf{b}:=\mathbf{a}\mathbf{b}$, then such a system
would satisfy many axioms of a conventional algebra over a field,
including distributivity of multiplication over addition; however,
scalar multiplication would not be homogeneous since for example $\left(\lambda\mathbf{a}\right)\odot\mathbf{b}=\exp\left(\ln\left(\lambda\mathbf{a}\right)\ln\mathbf{b}\right)=\exp\left(\ln\lambda\ln\mathbf{b}\right)\left(\mathbf{a}\odot\mathbf{b}\right)\neq\lambda\left(\mathbf{a}\odot\mathbf{b}\right)$.
It would be consistent with axioms stated in the original work
but not support the conclusion. In general, algebraic constructs that
result from dropping homogeneity of vector multiplication, $\left(\lambda\mathbf{a}\right)\odot\left(\mu\mathbf{b}\right)=\lambda\mu\left(\mathbf{a}\odot\mathbf{b}\right)$,
while maintaining distributivity of multiplication over addition,
have been investigated under the ``commutative hyperoperation''
label and proposed as early as \cite{Bennett1915}.}. We showed that probabilities are quadratic in amplitudes without
needing to know the dimensionality of the vector space. All along,
we followed a program that aimed at providing clarity of thought at
the interface of mathematics and physics: We separated axioms from
choices in an attempt to differentiate what properties of the model
derive from foundational assumptions, in contrast to properties that
trace to choices made by an outside experimenter in order to provide
answers for specific questions.

Curiously, the purely mathematical nature of composition algebras
over the reals prescribes the dimensionality of allowable amplitude
algebras. We found no grounds for excluding the two-dimensional split-complex
numbers from permissible amplitude algebras, as well as the four-dimensional
quaternions and split-quaternions. Quaternions do not occur in the
original work due to the fixed requirement of a two-dimensional configuration
space. To rule out the split-algebras in two and four dimensions,
there would have to be some triangle inequality constraint for $\sqrt{p}$
(e.g.~\cite{Baez2002}), possibly from convexity axioms on probabilities;
or alternatively, required existence of a nondegenerate trilinear
form in the amplitude algebra (triality; also \cite{Baez2002}). We
did not find a corresponding structure in the model from which such
could be traced.

While we weakened or eliminated several postulates from the original
work under investigation, the following algebraic properties still
have to be required ad-hoc, i.e., without tracing to the model or
the question for probabilities: commutativity of scalar multiplication
and continuity of the $p$-function. These seemingly small postulates
have far-reaching consequences: Allowing scalars to be noncommutative
would generalize the space of amplitudes from being a vector space
to merely a module. It would require clarification on how to generalize
the quadratic form $Q$ and its behavior under involution of the argument
(\ref{eq:QisQofInvolution}), and even more so, it would entirely
leave the scope of Jacobson's axioms classifying composition algebras.
Conversely, without continuity of the $p$-function, its exponential
behavior (\ref{eq:pFunctionIsHomogeneousOfDegreeAlpha}) and signomial
representation of the $p$-functions (\ref{eq:pAsHomogeneousSignomial})
could not be deduced. Not knowing whether $p\left(\lambda\mathbf{1}\right)=0$
infers $\lambda=0$ (\ref{eq:pLambda1eq0infersLambda0}) would leave
a gap in satisfying the field axiom for multiplicative inverses for
scalars. Not knowing whether $p\left(\lambda\mathbf{1}\right)=1$
infers $\lambda=\pm1$ (\ref{eq:pLambda1eq1infersLambda1}) would prevent
the conclusion that fully coarsening scalars must result in the unit
amplitude (\ref{eq:fullyCoarseningScalarsIsUnitAmplitude}). All of
these are required to show that the $p$-functions are indeed a quadratic
form.

As a final mathematical call-out, we motivated real-number representation
of the scalar field from a simplicity argument in section \ref{subsec:Scalar-field}
to keep the mathematical space of amplitudes as similar as possible
to the space of probabilities they ought to explain. It is nevertheless
a relatively strong choice. As hinted at with examples in that section,
weakening this choice could take the work in different directions.

From the physics point of view, we hope that our improved clarity
in treating this elementary building block of quantum reality provides
insight into existing follow-on work that considers composite systems
\cite{Goyal2014} and distinguishability \cite{Goyal2015} and allows
the model to be incorporated into new or existing work that reconstructs
spacetime, the light cone, and the dynamical interaction of quantum
systems\footnote{Prominent graph-based approaches that address these questions include
Causal Set theory, which offers powerful insights into proposed structures
of spacetime, and the intriguing spin networks of Loop Quantum Gravity.
For recent review articles, see e.g.~\cite{Rovelli2012,Surya2019,AB2021,Surya2025}.}.

\subsection{Histories matter for amplitudes?}

When we asked for probabilities of measurement outcomes in section
(\ref{sec:Probability}), we required that any experiment be independent
of the system's history prior to the first measurement. This concept
of closure was part of our question; hence, we required it from the
representation from which we expect our answer. In future work, we
could subtly modify our question, to still require that histories
of a system do not matter for \emph{probabilities} of a subsequent
measurement; however, there may exist further degrees of freedom in
the amplitude algebra that do not contribute to future probabilities.
These degrees of freedom may depend on a system's history and, with
that, no longer be subject to closure as defined in this paper. In
a way, this weaker form of closure would resemble an anthropocentric
stance, where properties of nature that are conceptually out of reach
for a human observer cannot be prepared for in an experiment, are
driven by the system's past, and consequently drive behavior of that
system that likewise remains out of human reach.

Regardless of whether the current paper's concept of (strong) closure
or the proposed weak closure would be assumed when formulating questions
to the system, the resulting answers for probabilities would remain
the same. That does not mean, however, that postulating weak closure
would therefore be inconsequential. To sketch a window of opportunity
for this ansatz, its outcome could be that the degrees of freedom
inaccessible to humans contribute to measurement outcomes in a way
that would make certain aspects of the outcome follow seemingly random
patterns. Those patterns could be irreconcilable with classical probability
yet follow well-defined probability distributions. Nature, by supposition,
would answer a question for probabilities by fitting its inaccessible
inner workings into the template provided by that very question. This
sketch traces observed properties from canonical Quantum Mechanics,
of course.

Under weak closure, it would not be allowable anymore to take an amplitude
of an existing path and prepend a new measurement to it, as this could
amount to altering the past of the existing path. Because histories
now matter, we would only be allowed to model further experiments
by \emph{appending} measurements to existing ones. In formulas, starting
with two measurement sequences $\mathcal{M}^{\mathrm{A}}$ and $\mathcal{M}^{\mathrm{B}}$
that are chained, $\mathcal{M}^{\mathrm{A}}\cdot\mathcal{M}^{\mathrm{B}}$,
we would only be allowed to chain a third sequence $\mathcal{M}^{\mathrm{C}}$
to the side of the expression that represents the future (say, to
the right) because chaining it to the side that represents the past
(the left) would correspond to altering history under weak closure.
Expressions like $\left(\mathcal{M}^{\mathrm{A}}\cdot\mathcal{M}^{\mathrm{B}}\right)\cdot\mathcal{M}^{\mathrm{C}}$
would remain allowable, but expressions from the mere regrouping of
the order of brackets, $\mathcal{M}^{\mathrm{A}}\overset{??}{\cdot}\left(\mathcal{M}^{\mathrm{B}}\cdot\mathcal{M}^{\mathrm{C}}\right)$,
could not be traced from the model anymore.

Algebraically, this permits algebras, where multiplication is not
necessarily associative. The reader readily recognizes the nonassociative
octonions $\mathbb{O}$ and split-octonions $\mathbb{O}^{\prime}$
to become permissible \cite{Okubo1995,Baez2002,ConwaySmithD2003} since associativity of multiplication
was the only property that ruled out these real composition algebras
in this paper. Just as for the complexes and quaternions, their $p$-function
models probability, which does remain invariant when modifying product
association. This, therefore, still satisfies the weak closure requirement
for probabilities.

While algebraic nonassociativity may feel inconvenient at first, it
is essentially no different than noncommutativity in the quaternions
$\mathbb{H}$, which likewise does not affect the $p$-function of
the product\footnote{Or noncommutativity in matrix multiplication, which does not affect
the determinant.}. Even more so, building an algebra from one-sided multiplication
only and letting products act on a separate element $\psi$ of that
algebra multiplicatively forms an \emph{associative} algebra by definition:
It simply forces all brackets a certain way. Using such one-sided
multiplication algebras built from octonions with real coefficients,
$\mathbb{O}$, with complex coefficients $\mathbb{C}\otimes\mathbb{O}$,
quaternionic coefficients $\mathbb{H}\otimes\mathbb{O}$, and iteratively
further ``tensoring'' these to form algebras like
$\mathbb{C}\otimes\mathbb{H}\otimes\mathbb{O}$ and larger, all form
Clifford algebras; and that separate element $\psi$ these act on
multiplicatively hence is identified as a spinor\footnote{By definition, a spinor can be understood as an element on
which a Clifford algebra acts.}. This construction has been explored in physics for use in modeling
symmetries in the Standard Model of particles since the late 20th
century \cite{Dixon1994,Dixon2014}. Significant recent progress towards
a deeper, more complete understanding of the various algebraic aspects,
possible realizations, degrees of freedom, consequences, and constraints
\cite{Furey2016,FuH2022-OneGen,FuH2022-SymmBr,Furey2023rmap1} led
to active research in the field. A recent concrete proposal for
complex-octonionic operators for the strong force \cite{RSC2024}
is used to demonstrate core properties of nature in a Feynman
diagram setting \cite{RC2025}. Curiously, these endeavors use the
same algebras that would become permissible in our model here when
allowing weak closure for answering an experimenter's question.

Prescribing the order of multiplication also means that all algebraic
properties must now only be satisfied in one-sided multiplication,
namely, for the side from which it is allowed to multiply a priori.
Sometimes called ``(left-/right-)multiplication algebras,''\cite{Schafer1966}
there are initial discoveries of normed multiplication algebras that
permit a linear, homogeneous multiplication from one side but are
nondistributive and nonhomogeneous when multiplying from the other
\cite{SmithJDH1995,ConwaySmithD2003,SmithW2004}. A complete mathematical
classification is subject to ongoing investigation.

In nonassociative multiplication algebras, there is no concept of
inserting unit amplitudes $\mathbf{1}$ through chaining at different
places in a measurement sequence. Such an operation would not be allowed,
as you may multiply only from one side. With that, the existence of
a multiplicative unit \emph{amplitude} $\mathbf{1}$ in the algebra
becomes optional, permitting so-called nonunital algebras. This is
distinct from the existence of a unit \emph{scalar}, $1$, which can
always be modeled into any sequence through redundancy from duplicated
measurements. In our work, scalars are separate from amplitudes by
definition, which is consistent with the new freedom. Nonunital algebras
over the reals include the eight-dimensional Okubo algebra \cite{Okubo1995,SmithVojtech2022}.
In the reals, normed nonunital algebras with product composition still
permit triality \cite{Elduque1996,Elduque2000,Elduque2022,KO2015,KO2016}.
This key property of algebras that would become permissible in our
model, namely, of generally not associative or unital multiplication
algebras with triality\footnote{Classification of these algebras is an open question in mathematics
as well.}, is likewise explored in contemporary algebraic physics research
\cite{Hughes2016,Koepl2023aut,MCZ2023,FuH2024}.

\subsection{Do\emph{ all} histories matter?}

Assuming that all histories matter when calculating amplitudes, an
earlier graph-based model was captured in \cite{Furey2016} (section
2) that postulates for nodes from all elements in the lower set of
a partially ordered set (from the ``poset ideal'') to contribute
to an algebraic expression. When asking for an observation, that expression
will be evaluated; otherwise, when left untouched by an observer,
the expression will keep accumulating nodes indefinitely as the system
evolves. This extreme concept of ``all histories matter'' supports
an intuition towards amplitude representation in the form of algebraic
ideals, i.e., algebraic subspaces that absorb general amplitudes into
their respective subspace under multiplication yet are closed under
amplitude addition. Under some conservation assumptions, suitable
algebras in which these ideals possess symmetries that resemble the
ones observed in the Standard Model of particles in physics are found
in tensored division algebras over the reals, $\left(\mathbb{R}\otimes\right)\mathbb{C}\otimes\mathbb{H}\otimes\mathbb{O}$
\cite{Furey2010} -- yet again.

The occurrence of related algebraic structures in different contexts,
now even tracing from different model assumptions, leads one to wonder
about a common provenance. How could our current ``histories don't
matter'' postulate be reconcilable with an ``all histories matter''
axiom? Closing on an inspirational note, when weakening closure from
``histories don't matter'' to include ``\ldots{} for probabilities
only,'' to become the anthropocentric consequence of an observer
approaching nature with the very question for probabilities, it is
conceivable that our model here describes the behavior of quantum
nature under observation\footnote{Quoting \cite{Goyal2012}: Quantum Mechanics may be ``{[}not{]} a
description of reality in itself {[}but{]} a description of reality
as experienced by an agent''.}; whereas ``all histories matter'' approaches like superdeterminism\cite{AHHP2024}
describe the inner workings of nature when left unobserved\footnote{For a sketch of an agency-free model of Quantum Field Theory, on algebras
permissible when extending our current work, see e.g.~ \cite{Furey2023osmu}.}.

\backmatter

\bmhead{Acknowledgements}

We would like to extend our sincere thanks to Nichol Furey, Bernd
Henschenmacher, John Huerta, and John A.~Shuster for their helpful
discussion, pointers, corrections, and mentoring, as well as to the
organizers and attendees of the ``Mile High Conferences on Nonassociative
Mathematics'' (2009, 2013, and 2017) at the University of Denver.

\section*{Declarations}

M.~Habeck gratefully acknowledges funding by the Carl Zeiss Stiftung within the program ``CZS Stiftungsprofessuren.'' The authors have no conflicts of interest to declare that are relevant to the content of this article.

\end{document}